%% file: tit-2019.tex
\documentclass[journal, english, onecolumn]{IEEEtran}
\input{preamble}

\title{Tracking an Auto-Regressive Process with Limited Communication per Unit Time}  
\author{ Rooji~Jinan
\quad \and Parimal~Parag
\quad \and Himanshu~Tyagi

\thanks{The authors are with the Department of Electrical
  Communication Engineering, Indian Institute of Science, Bangalore
  560012, India.  Email: \{roojijinan, parimal, htyagi\}@iisc.ac.in.}
}
\begin{document}
\maketitle
\input{abstract}

\input{intro-TIT}
\input{setup}

\input{successive-update}
\input{results}

\input{analysis_SU}
\input{achievability}

\input{converse}

\input{discussion}
\bibliographystyle{IEEEtranS}
\bibliography{tit-2019}
\end{document}

%% file: preamble.tex
\usepackage[cmex10]{amsmath}
\usepackage[mathscr]{euscript}
\usepackage[utf8]{inputenc}
\usepackage{%
	amssymb,%
	amsthm,%
	babel,%
	bbm,%
	cite,%
	mathtools,%
	pgf,%
	pgfplots,%
	tikz,
	caption,
	subcaption%
	}
\usepgflibrary{shapes}
\usetikzlibrary{%
  arrows,%
  backgrounds,%
  calc,%
  calendar,%calendar
  chains,%
  decorations,%decorations
  decorations.pathmorphing,% /pgf/decoration/random steps | erste Graphik
  fit,%
  matrix,%
  mindmap,%mindmap
  petri,%
  positioning,% wg. " of "
  scopes,%
  shadings,%
  shadows,%
  shapes,%
  shapes.arrows,%
  shapes.misc,% wg. rounded rectangle
  shapes.symbols,%
  snakes,%snakes 
}
\definecolor{UBCblue1}{RGB}{81, 191, 218}
\definecolor{UBCblue2}{RGB}{153, 219, 236}
\definecolor{UBCblue3}{RGB}{80, 180, 200}
\tikzset{%
	sbStyleBloc/.append style = {align = center}}
\tikzset{
    block/.style = {draw, rectangle, 
        minimum height=1.5cm, 
        minimum width=2.5cm,fill=blue!10},
    input/.style = {coordinate,node distance=2cm},
    output/.style = {coordinate,node distance=4.5cm},
    arrow/.style={draw, -latex,node distance=2cm},
    pinstyle/.style = {pin edge={latex-, black,node distance=4cm}},
    sum/.style = {draw, circle, node distance=2cm}
}
\usepackage{verbatim}
\pgfplotsset{compat=1.12}
\theoremstyle{plain}
\newtheorem{thm}{Theorem}%[section]
\newtheorem{lem}[thm]{Lemma}

\theoremstyle{definition}
\newtheorem{defn}{Definition}

\newtheorem{exmp}{Example}

\theoremstyle{remark}

\DeclarePairedDelimiter\abs{\lvert}{\rvert}%
\DeclarePairedDelimiter\norm{\lVert}{\rVert}%
\DeclarePairedDelimiter\set{\lbrace}{\rbrace}%
\DeclarePairedDelimiter\inner{\langle}{\rangle}%

\renewcommand{\ge}{\geqslant}
\renewcommand{\le}{\leqslant}

\newcommand{\R}{\mathbb{R}}
\newcommand{\N}{\mathbb{N}}
\newcommand{\E}{\mathbb{E}}

\newcommand{\Z}{\mathbb{Z}}

\newcommand{\cC}{\mathcal{C}} 
 \newcommand{\cN}{\mathcal{N}}

\newcommand{\tI}{\tilde{I}}

\newcommand{\eq}[1]{\begin{align*}#1\end{align*}}
\newcommand{\EQ}[1]{\begin{equation*}#1\end{equation*}}
\newcommand{\eqn}[1]{\begin{align}#1\end{align}}
\newcommand{\EQN}[1]{\begin{equation}#1\end{equation}}

\newcommand{\bEE}[1]{\mathbb{E}\left[#1\right]}
\newcommand{\bPr}[1]{\mathbb{P}\left(#1\right)}

\newcommand{\eps}{\varepsilon}
\renewcommand{\set}[1]{\left\lbrace#1\right\rbrace}
\newcommand{\SetIn}[1]{\mathbbm{1}_{#1}}

\newcommand{\ieeeproof}[1]{\begin{IEEEproof}#1\end{IEEEproof}}

\DeclareMathOperator*{\argmin}{arg\,min}

%% file: abstract.tex
\begin{abstract} 
  Samples from a high-dimensional AR[1] process are 
  observed by a sender which can communicate only finitely many
bits per unit time to a receiver.
  The receiver seeks
  to form an estimate of the process value at every time instant in real-time.
We consider a time-slotted communication model in a
slow-sampling regime where multiple communication slots occur between
two sampling instants.
We propose a {\em successive update} scheme
which uses communication between sampling instants to refine estimates
of the latest sample and study the following question: Is it better to 
collect communication of multiple slots to send better refined estimates,
making the receiver wait more for every refinement,
or
to be {\em fast but loose} and send new information in every
communication opportunity?  We show that the fast but loose successive
update scheme with ideal spherical codes
is universally optimal asymptotically for a large dimension. However,
most practical quantization codes for fixed dimensions do not meet the ideal performance 
required for this optimality, and they typically will have a bias in the form of a fixed
additive error. Interestingly, our analysis shows that 
the fast but loose scheme is not an optimal choice 
in the presence of
such errors, and a judiciously
chosen frequency of updates outperforms it.
\end{abstract}

%% file: intro-TIT.tex
\section{Introduction}
\label{sec:Intro}
We consider the setting of real-time decision systems based on
remotely sensed observations.  In this setting, the decision maker
needs to track the remote observations with high precision and in a
timely manner.  These are competing requirements, since high precision
tracking will require larger number of bits to be communicated,
resulting in larger transmission delay and increased staleness of information.
Towards this larger goal, we study the following problem.

Consider a discrete time first-order auto-regressive (AR[1]) process $X_t\in\R^n$, $t\geq 0$. A sensor
draws a sample from this process, periodically once every $s$ time-slots. In each of these time-slots, the sensor can send $nR$ bits to a center. The center seeks to form an estimate $\hat{X}_t$ of $X_t$ at time $t$, with small mean square error (MSE). Specifically, we are interested in minimizing the time-averaged error $\sum_{t=1}^T\E{\norm{X_t-\hat{X}_t}_2^2}/T$ to enable timely and accurate tracking of $X_t$.

We propose and study a {\it successive update} scheme where the encoder
computes the error in the estimate of the latest sample at the decoder and sends its quantized value to the decoder. The decoder adds this value to its previous estimate to update the estimate of the latest sample, and uses it to estimate the current value using a linear predictor. We instantiate this scheme with a general gain-shape quantizer for error-quantization.

Note that we can send this update several times between two sampling instances. In particular, our interest in comparing a {\em fast but loose} scheme where an update is sent every slot or a slower update every $p$ communication slots. The latter allows the encoder to use more bits for the update, but the decoder will need to wait longer. We analyze this scheme for a universal setting and show that the fast but loose successive update scheme, used with an appropriately selected quantizer, is optimal asymptotic.

To show this optimality, we use a random construction for the quantizer, based on the spherical code given in~\cite{Wyner1967, Lapidoth1997TIT}. Roughly speaking, this ideal quantizer $Q$ yields
\[
\E{\norm{y-Q(y)}_2^2}\leq \norm{y}_2^22^{-2R},
\]
for every $y$ uniformly bounded. However, in practice, at finite $n$,
such quantizers need not exist. Most practical vector quantizers have
an extra additive error, $i.e.$, the error bound takes the form
\[
\E{\norm{y-Q(y)}_2^2}\leq \norm{y}_2^2\theta + n\eps^2.
\]
We present
our analysis for such general quantizers. Interestingly, for such a
quantizer (which is all we have at a finite $n$), the optimal choice
of $p$ can differ from $1$. Our analysis provides a theoretically sound
guideline for choosing the frequency of updates $1/p$ for practical
quantizers. 

%\subsection{Related works}
Our work relates to a large body of literature ranging from real-time
compression to control and estimation over networks. 
The structure of real-time encoders for source coding has been studied
in~\cite{Witsenhausen1979, Teneketzis2006TIT, Mahajan2007ITW,
  Walrand1983TIT, YukselS2012TIT, Linder2014TIT, Wood2017TIT}.  The
general structure of real-time encoders for Markov sources is studied
for communication over error-free channels in~\cite{Witsenhausen1979}
and over noisy channels in~\cite{Teneketzis2006TIT, Mahajan2007ITW}.
A similar structural result for the optimal encoders and decoders
which are restricted to be causal is given in~\cite{Walrand1983TIT}.
Furthermore, structural results in the context of optimal zero-delay
coding of correlated sources are available in~\cite{YukselS2012TIT,
  Wood2017TIT, Linder2014TIT}. 
  The setup in all these works are different from the problem we consider
  and the results do not extend to our problem.

The problems of remote estimation under communication
constraints of various kinds have been studied in~\cite{Wong1997TAC,
  NairEvans97, NairEvans98, Dokuchaev1999CDC, Smith2003TAC, Matveev2003TAC}.
This line of work proposes several Kalman-like recursive estimation algorithms and evaluates their performances. In a related thread,~\cite{Lipsa2011TAC, Chakravorty2017TAC,Basar2013TAC}
study 
remote estimation under communication constraints and other related constraints 
using
tools from dynamic programming and stochastic control. 
{However, in all these works the role of channel delay is slightly different from that in our setting.
  Furthermore, the specific problem of choice of quantizer we consider has not been looked at.}
More recently,~\cite{Sun2017ISIT} studied remote estimation of Wiener
process for channels with random delays and proves that the optimal
sampling policy is a threshold based policy. 
{This work, like some of the works cited above,
assumes real-valued transmissions and does
not take into account any quantization effects}.

{ In more information theoretic settings, 
sequential coding for individual sequences under delay constraints was 
studied
in~\cite{Linder2001TIT,Weissman2002TIT,Weissman2001TIT,Matloub2006TIT}. 
Closer to our work, the causal (non-anticipatory) rate-distortion function for a stochastic process
goes back to early works~\cite{GorPin73,GorPin74}. Recent
works~\cite{StavrouKC14,StavrouOC18, StavrouCCL18} consider the
specific case of auto-regressive and Gauss-Markov processes and
use the general formula in these early works to establish asymptotic
optimality of simpler information structure for the encoders (optimal decoder
structure is straightforward).  
We note that related formulations have been studied for simple
settings of two or three iterations in~\cite{Berger2000TIT,MaIshwar11}
where interesting encoder structures for using previous communication for next
sample as well emerge. Although some of these works propose specific
optimal schemes as well, the key results in this
line of research provide an expression for the rate-distortion
function as an optimization problem, solving which will provide
guidelines for a concrete scheme. 
In contrast, motivated by problems of estimation over an erasure
channel,~\cite{KhinaKKH19} provides an asymptotically optimal scheme
that roughly uses Gaussian codebooks to quantize the innovation errors
between the encoder's observation and the decoder's estimation. Our
work is closest to~\cite{KhinaKKH19}, but differs in an important
aspect from all the works mentioned in this research thread: We
take transmission delays into account. In particular, in our
formulation the decoder estimation and communication delays are on the same
time-scales. The decoder must provide an estimate at every time
instant, and a longer codeword will result in a longer delay for the
decoder to get complete information.}

{Nonetheless, our converse bound is derived using similar methods as~\cite{Khina2017ITW}. 
Even the achievability part of our proof draws
from~\cite{Khina2017ITW}, but there is technical caveat.  Note that
after the first round of quantization, the error vector need not be
Gaussian, and the analysis in~\cite{Khina2017ITW} can only be applied
after showing a closeness of the error vector distribution to Gaussian
in the Wasserstein distance of order $2$. While the original
proof~\cite{Khina2017ITW} overlooks this technical point, this gap can
be filled using a recent result from~\cite{KipnisReeves19} if
spherical codes are used. But we follow an alternative approach and
show a direct analysis using vector quantizers.}

{In addition, there is a large body of work on control problems over
rate-limited communication channels ($cf.$~\cite{Delchamps89, Borkar1997,
  Wong1999TAC,  Nair2000CDC, Liberzon2003TAC, You2010TAC,
  Yuksel2010TAC, Yuksel2012TAC, Yuksel2012TIT, Yuksel2016SIAM}).
This line of work implicitly requires handling of communication delays
in construction of estimators. However, the simple formulation we have
seems to be missing, and results in this long line of work do not
resolve the questions we raise.} 

%% While these works assume state estimation over perfect channels, the authors
%% of~\cite{Smith2003TAC, Matveev2003TAC} consider the state estimation
%% of linear systems over lossy channels. Again, these works are related
%% to our work, but the formulation is different from ours and does not
%% provide much insights for the questions we raise. 
%Also, here the authors uses a linear measurement model and discusses the optimal filtering techniques under various settings. 
	
%% We omit the details as
%% the requirement for stability and optimal control is different from
%% our goal of minimizing the estimation error.

Our main contributions in this paper are as follows: We present a heuristically appealing encoder
structure which we show to be optimal asymptotically in the dimension
of the observation. Specifically, we propose to send successive
updates that refines the estimates of the latest sample at the
decoder. It is important to note that we quantize the estimation
  error at the decoder, instead of quantizing the innovation sequence
  formed at the encoder.  Although the optimal MMSE decoder involves
  taking conditional expectation, we use a simple decoder which
uses a simple linear structure. Yet, we show that this decoder is
asymptotically optimal.
Then, we instantiate this general scheme with
spherical codes for quantizers to obtain a universal scheme.
In particular, we consider general gain-shape quantizers and develop a
framework to analyze their performance. One interesting result we
present shows that the tradeoff between the accuracy and frequency of
the updates must be carefully balanced based on the ``bias'' (additive
error) in the
quantizer used.

We present our problem formulation in the next section.
Section~\ref{sec:successive_update} presents a discussion on the
our achievability scheme followed
by the main results in the subsequent section. 
Section~\ref{sec:analysis_su} provides a detailed analysis of our
scheme, which we further build-on in
Section~\ref{sec:asymptotic_achievability} to get our asymptotic
achievability results. 
We prove our converse bound in Section~\ref{sec:converse}, 
and conclude with a discussion on extensions of
our result in the final section. 

%% file: setup.tex
%%%%%%%%%%%%%%%%%%%%%%%%%%%%%%%%%%%%%%%%
\section{Problem Formulation}
\label{sec:prblm_form}
%%%%%%%%%%%%%%%%%%%%%%%%%%%%%%%%%%%%%%%%
We begin by providing a formal description of our problem; different
components of the model are presented in separate sections.
Throughout the remainder of this paper, the set of real numbers is
denoted by $\R$, the $n$-dimensional Euclidean space is denoted by
$\R^n$ and the associated Euclidean norm by $\norm{\cdot}_2$, the set of
positive integers is denoted by $\N$, the set of non-negative integers
is denoted by $\Z_+$, the set of continuous positive integers until
$m$ is denoted by $[m] \triangleq \set{1, \dots, m}$, and an identity
matrix of size $n \times n$ is denoted by $I_n$.

\subsection{Observed random process and its sampling}
For $\alpha\in (0,1)$, we consider a discrete time auto-regressive
process of order 1 (AR[1] process) in $\R^n$, \EQN{
\label{eqn:process}
X_{t} = \alpha X_{t-1} + \xi_t, \quad t\ge 0, } where $(\xi_t \in
\R^n, t \ge 1)$ is an independent and identically distributed
(\emph{i.i.d.}) random sequence with zero mean and covariance matrix
$\sigma^2(1-\alpha^2)I_n$.  For simplicity, we assume that $X_0 \in
\R^n$ is a zero mean random variable with covariance matrix
$\sigma^2I_n$.  This implies that the variance of $X_t \in \R^n$ is
$\sigma^2I_n$ for all $t \ge 0$.  In addition, we assume that
$\norm{X_t}_2$ has a bounded fourth moment at all times $t \ge 0$.
Specifically, let $\kappa>0$ satisfy
%be the per-dimension uniform bound on the fourth moment, i.e.
\[
\sup_{k \in \Z_+}\frac{1}{n} \sqrt{\E\norm{X_k}_2^4} \le \kappa.
\]
It is clear that $X = (X_t \in \R^n, t \ge 0)$ is a Markov process.
We denote the set of processes $X$ satisfying the assumptions above by $\mathbb{X}_n$
and the class of all such processes for different choices of dimension $n$ as $\mathbb{X}$.

This discrete time process is sub-sampled periodically at {\em
  sampling frequency} $1/s$, for some $s\in \N$, to obtain samples
$(X_{ks} \in \R^n, k \ge 0)$.

\subsection{Encoder description}
The sampled process $(X_{ks}, k \ge 0)$ is passed to an {\em encoder}
which converts it to a bit stream.  The encoder operates in {\em
  real-time} and sends $nRs$ bits between any two sampling instants.
Specifically, the encoder is given by a sequence of mappings
$(\phi_t)_{t\geq 0}$, where the mapping at any discrete time $t = ks$
is denoted by 
\EQ{ 
\phi_t:\R^{n(k+1)} \to \set{0,1}^{nRs}.  
} 
The encoder output at time $t=ks$ is denoted by the codeword $C_t
\triangleq \phi_t(X_0, X_s, \dots, X_{ks})$.  We represent this codeword
by an $s$-length sequence of binary strings $C_t = (C_{t,0}, \dots,
C_{t,s-1})$, where each term $C_{t,i}$ takes values in
$\set{0,1}^{nR}$. For $t=ks$ and $0\leq i \leq s-1$, we can view the
binary string $C_{t,i}$ as the communication sent at time $t+i$. We
elaborate on the communication channel next.

%The channel can send $nR$ bits per transmission. 
%\red{Specifically, denoting by $\cX$ the source alphabet}
%$\R^n$, at time instant $t=ks$, the encoder uses a mapping $g_t:
%\cX^{k+1}\to \{0,1\}^{nRs}$ to get the codeword $C_t=g_t(X_0, X_{s},
%\dots, X_{ks})$. The codeword $C_t=(C_{t,1}, \dots, C_{t,s})$ where
%each $C_{t,i}$ is an $nR$ bit vector. 

\subsection{Communication channel}
The output bit-stream of the encoder is sent to the receiver via an
error-free communication channel. Specifically, we assume slotted
transmission with synchronization where in each slot the transmitter
sends $nR$ bits of communication error-free.  That is, we are allowed
to send $R$ bits per dimension, per time slot.  Note that there is a
delay of $1$ time-unit (corresponding to one slot) in transmission of
each $nR$ bits.  Therefore, the vector $C_{ks,i}$ of $nR$ bits
transmitted at time $ks+i$ is received at time instant $ks+i+1$ for $0
\le i \le s-1$.  Throughout we use the notation $I_k \triangleq
\set{ks, \dots, (k+1)s-1}$ and $\tI_k=I_k+1=\set{ks+1, \dots,
  (k+1)s}$, respectively, for the set of transmit and receive times
for the strings $C_{ks,i}$, $0\leq i \leq s-1$.
%and the set $\tI_k=I_k+1$ corresponds to the receive times for $C_{ks,i}$. 

\subsection{Decoder description} 
We describe the operation of the receiver at time $t \in I_k$, for some
$k \in \N$, such that $i = t-ks \in \set{0, \dots, s-1}$.
Upon receiving the codewords $C_s, C_{2s}, ...,C_{(k-1)s}$ 
and the partial codeword $(C_{ks,0},..., C_{ks,i-1})$ at time $t=ks+i$, 
the {\em decoder} estimates the current-state $X_t$ of the process using the estimator mapping
\EQ{
\psi_t:\{0,1\}^{nRt} \to \R^n.  
}
We denote the overall communication received by the decoder until time instant\footnote{The time index $t-1$ in $C^{t-1}$ corresponds to the transmission time of the codewords, whereby the communication received till time $t$ is denoted by $C^{t-1}$.} $t$ by $C^{t-1}$.
%With a slight abuse of notation, we denote by
%$C^t$ the overall communication received until time instant $t$.
Further, we denote by $\hat{X}_{t|t}$ the real-time causal estimate $\psi_t(C^{t-1})$ of $X_t$ formed
at the decoder at time $t$. 
Thus, the overall real-time causal estimation scheme is
described by the mappings $(\phi_t,\psi_t)_{t\ge 0}$. 
It is important to note that the communication available to the decoder at time $t \in I_k$ can only depend on 
samples $X_\ell$ up to time $\ell \le ks$.
%t-(t\mod s) = s\lfloor t/s\rfloor$. 
As a convention, we assume that $\hat{X}_{0|0}= 0$. 

\subsection{Performance metrics}
We call the encoder-decoder mapping sequence $(\phi, \psi)=(\phi_t, \psi_t)_{t \ge 0}$ a tracking code of rate $R$ and sampling period $s$. The tracking error of our tracking code at time $t$ for process $X$ is measured by the mean squared error (MSE) per dimension given by 
\EQ{
D_t(\phi,\psi, X) \triangleq \frac{1}{n}\E\norm{X_{t} - \hat{X}_{t\vert t}}_2^2.
}
Our goal is to design $(\phi,\psi)$ with low average tracking error $\overline{D}_T(\phi,\psi,X)$ given by 
\[
\overline{D}_T(\phi,\psi,X) \triangleq \frac 1 {T} \sum_{t=0}^{T-1} D_t(\phi,\psi,X).
\]
For technical reasons, we restrict to a finite time horizon setting. For the most part, the time horizon $T$ will remain fixed and will be omitted from the notation. Instead of working with the mean-square error, a more convenient parameterization for us will be that of accuracy, given by
\[
\delta^T(\phi, \psi, X)= 1-\frac{\overline{D}_T(\phi,\psi,X)}{\sigma^2}.
\]

\begin{defn}[Maxmin tracking accuracy]
\label{def:accuracy}
The worst-case tracking accuracy for $\mathbb{X}_n$ attained by a tracking code $(\phi,\psi)$ is given by 
\[
\delta^T(\phi,\psi,\mathbb{X}_n) =\inf_{X\in \mathbb{X}_n} \delta^T(\phi, \psi, X).
\]
The maxmin tracking accuracy for $\mathbb{X}_n$ at rate $R$ and sampling period $s$ is given by
\[
\delta_n^T(R, s, ,\mathbb{X}_n)=\sup_{(\phi, \psi)}\delta^T(\phi,\psi,\mathbb{X}_n),
\]
where the supremum is over all tracking codes $(\phi, \psi)$. 
\end{defn}
The maxmin tracking accuracy $\delta_n^T(R, s,\mathbb{X}_n)$ is the fundamental quantity of interest for us.
Recall that $n$ denotes the dimension of the observations in $X_t$ for $X\in \mathbb{X}_n$ and $T$ the time horizon.  However, we will only characterize $\delta_n^T(R,s,\mathbb{X}_n)$ asymptotically in $n$ and $T$. Specifically, we define the asymptotic maxmin tracking accuracy as
\[
\delta^*(R, s, \mathbb{X}) = \limsup_{T\to \infty}\limsup_{n\to \infty} \delta_n^T(R,s,\mathbb{X}_n). 
\]
We will provide a characterization of $\delta^*(R, s, \mathbb{X})$ and
present a sequence of tracking codes that attains it. In fact, the tracking code we use
is an instantiation of our successive update scheme, which we describe in the next section. It is important to note that our results may not hold if we switch the order of limits above: We need very large codeword lengths depending on a fixed finite time horizon $T$.

%% file: successive-update.tex
%%%%%%%%%%%%%%%%%%%%%%%%%%%%%%%%%%%%%%%%
\section{The Successive Update Scheme} 
\label{sec:successive_update}
%%%%%%%%%%%%%%%%%%%%%%%%%%%%%%%%%%%%%%%%
In this section, we present our main contribution in this paper, namely the {\em Successive Update} tracking code. Before we describe the scheme completely, we present its different components.
In every communication slot, the transmitter gets an opportunity to
send $nR$ bits. The transmitter may use it to send any information
about a previously seen sample. There are various options for the
encoder. For instance, it may use the current slot to send some
information about a sample it had seen earlier. Or it may use all the
slots between two sampling instants to send a quantized version of the latest sample.
Interestingly, it will be seen (quite straightforwardly) that there are not so many options for the decoder; it gets roughly fixed once the encoder
is chosen.
%%%%%%%%%%%%%%%%%%%%%%%%%%%%%%%%%%%%%%%%
\subsection{Decoder structure}
%As discussed in Section~\ref{sec:prblm_form}, 
%at each time $t \in I_k$, the encoder transmits $nR$ bits describing the samples observed so far. 
Once the quantized information is sent by the transmitter, at the
receiver end, the decoder estimates the state $X_{t}$, using the
codewords received until time $t$.
Since we are interested in forming estimates with small MSE, the decoder simply forms the 
minimum mean square error (MMSE) estimate using all the observations till that point. Specifically,
for $t\geq u$, 
denoting by $\tilde{X}_{u|t}$ the MMSE estimate $X_{u}$ formed by the communication $C^{t-1}$ received before
time $t$, we know ($cf.$~\cite{Poor1994})
\EQ{
\tilde{X}_{u|t} = \E[X_u| C^{t-1}].
}
The following result presents a simple structure for $\tilde{X}_{u|t}$ for our AR[1] model.
\begin{lem}[MMSE Structure] 
\label{lem:SimpleEstimate}
The MMSE estimates $\tilde{X}_{t|t}$ and $\tilde{X}_{t-i|t}$, respectively, of samples $X_{t}$ and $X_{t-i}$ at any time $t \in I_k$ and $i = t-ks$ using 
communication $C^{t-1}$ are related as
\EQ{
\tilde{X}_{t|t} = \alpha^i \tilde{X}_{t-i|t} = \alpha^i\E[X_{ks}|C^{t-1}]. 
}
\end{lem}
\ieeeproof{
Recalling the notation $I_k$, we can represent $t \in I_k$ as $t =
ks+i$ for $0\leq i\leq s-1$.  From the evolution of the AR[1] process,
for $1\leq i\leq s-1$, the sample $X_{ks+i}$ can be expressed in terms of the previous sample
$X_{ks}$ as
\EQN{
\label{recursion_for_X_(ks+i)_1}
X_{ks+i} = \alpha^{i}X_{ks}+ \sum_{j=1}^{i}\alpha^{i-j}\xi_{ks+j}, 
}
where the innovation sequence $(\xi_{ks+j}: j \ge 1)$ is independent of process samples $(X_0, \dots, X_{ks})$. 
By our specification, the historical observations $C^{t-1}$ at the
receiver depend only on the process evolution until time $ks$, namely $C^{t-1}$ is
independent of $(\xi_{ks+j}: j \ge 1)$ conditioned on $(X_0, \dots, X_{ks})$.
In particular, $\E[\xi_{ks+j}|C^{t-1}]=0$ for every $j\geq 1$. Thus, 
taking conditional expectation on both sides of~\eqref{recursion_for_X_(ks+i)_1}, we get
\EQ{
\tilde{X}_{ks+i|ks+i} = \E[X_{ks+i}|C^{t-1}] = \alpha^{i} \E[X_{ks}|C^{t-1}] = \alpha^{i}\tilde{X}_{ks|t},
}
since $\tilde{X}_{ks|t}=\E[X_{ks}|C^{t-1}]$. 
}
{Therefore, the optimal strategy for the decoder is to use the communication sent to form an estimate for the latest sample and then scale it to form the estimate of the state at the current time instant.}
%%%%%%%%%%%%%%%%%%%%%%%%%%%%%%%%%%%%%%%%
\subsection{Encoder structure: Refining the error successively}
The structure of the decoder exposed in Lemma~\ref{lem:SimpleEstimate} gives an important insight for encoder design: The communication sent between two sampling instants is used only to form estimates of the latest sample. 
In particular, the communication $C_{ks+1, i}$ transmitted at time $t=ks+i$ must be chosen to refine the previous estimate from
$\E[X_{ks}|C_0, \dots, C_{ks},C_{ks+1,0}, \dots, C_{ks+1, i-1}]$ to 
$\E[X_{ks}|C_0, \dots, C_{ks},C_{ks+1,0}, \dots, C_{ks+1, i-1}, C_{ks+1,i}]$.
This principle can be applied (as a heuristic) for any other form of the estimate as follows.
Let $\hat{X}_{ks|t}$ denote the estimate for $X_{ks}$ formed at the receiver at time $t$ (which need not be the MMSE estimate $\tilde{X}_{ks|t}$). Our encoder computes the error in the receiver estimate of the last process sample at each time instant $t$. Denoting
the error at time $t \in I_k$ by
$Y_{t} \triangleq X_{ks} - \hat{X}_{ks \vert t}$, 
the encoder quantizes this error $Y_t$ and sends it as communication $C_{ks+1, i}$.

Simply speaking, our encoder computes and quantizes the error in the
current estimate of the last sample at the decoder, and sends it to
the decoder to enable the refinement of the estimate in the next time
slot. While we have not been able to establish optimality of this
encoder structure, our results will show its optimality
asymptotically, in the limit as the dimension $n$ goes to infinity.

Even within this structural simplification, a very interesting
question remains. Since the process is sampled once in $s$ time slots, 
we have, potentially, $nRs$ bits to encode the latest sample. 
At any time $t \in \tI_k$, the receiver has access to $(C_0, \dots, C_{(k-1)s})$ 
and the partial codewords $(C_{ks,0}, \dots, C_{ks,i-1})$ for $i = t-ks$. 
A simple approach for the encoder is to use the complete codeword to express the latest sample 
and the decoder can ignore the partial codewords. This approach will result in slow but very accurate
updates of the sample estimates. An alternative 
 fast but loose approach will send $nR$ quantizer codewords to refine estimates in every communication slot.
Should we prefer fast but loose estimates or slow but accurate ones?
Our results will shed light on this conundrum.

%%%%%%%%%%%%%%%%%%%%%%%%%%%%%%%%%%%%%%%%
\subsection{The choice of quantizers}
In our description of the encoder structure above, we did not specify a key design choice, namely the choice of the quantizer. We will restrict to using the same quantizer to quantize the error in each round of communication. The precision of this quantizer will depend on whether we choose a fast but loose paradigm or a slow but accurate one. However, the overall structure will remain the same. Roughly speaking, we allow any gain-shape~\cite{gersho2012vector} quantizer which separately sends the quantized value of the gain $\norm{y}_2$ and the shape $y/\norm{y}_2$ for input $y$. Formally, we use the following abstraction. 

%We also associate an additive error factor $\eps$  with the quantizer.
\begin{defn}[$(\theta, \eps)$-quantizer family]
  \label{def:quantizer}
  Fix $0<M < \infty$. For $0\leq \theta\leq 1$ and $0\leq \eps$, a
  quantizer $Q$ with dynamic range $M$ 
specified by a mapping $Q: \R^n \to \{0,1\}^{nR}$
  constitutes an $nR$ bit {\em $(\theta, \eps)$-quantizer} if for
  every vector $y \in \R^n$ such that $\norm{y}_2^2 \le nM^2$,
  we have \EQ{ \E\norm{y - Q(y)}_2^2 \le \norm{y}_2^2
    \theta(R) + n\eps^2.  } Further, for a mapping $\theta: \R_+ \to
  [0,1]$, which is a decreasing function of rate $R$, a family of
  quantizers $Q=\{Q_R: R>0\}$ constitutes an {\em $(\theta,
    \eps)$-quantizer family} if for every $R$ the quantizer $Q_R$
  constitutes an $nR$ bit $(\theta(R), \eps)$-quantizer.
\end{defn}
%\begin{defn}[$(\theta, \eps)$-quantizer family]
%  \label{def:quantizer}
%  Fix $0<M < \infty$. For $0\leq \theta\leq 1$ and $0\leq \eps$, a
%  quantizer $Q$ specified by a mapping $Q: \R^n \to \{0,1\}^{nR}$
%  constitutes an $nR$ bit {\em $(\theta, \eps)$-quantizer} if for
%  every vector $y \in \R^n$ such that $\norm{y}_2^2 \le nM^2$,
%  we have \EQ{ \E\norm{y - Q(y)}_2^2 \le \norm{y}_2^2
%    \theta(R) + n\eps^2.  } Further, for a mapping $\theta: \R_+ \to
%  [0,1]$, which is a decreasing function of rate $R$, a family of
%  quantizers $Q=\{Q_R: R>0\}$ constitutes an {\em $(\theta,
%    \eps)$-quantizer family} if for every $R$ the quantizer $Q_R$
%  constitutes an $nR$ bit $(\theta(R), \eps)$-quantizer.
%\end{defn}
%Here, it is to be noted that expectation is taken with respect to the
%randomness in the quantizer.  
The expectation in the previous definition is taken with respect to the randomness in the quantizer,
which is assumed to be shared between the encoder and the decoder for
simplicity. 
The parameter $M$, termed the {\it dynamic range} of the quantizer,
specifies the domain of the 
quantizer. 
%The parameter $M$ in the description above specifies the domain of our
%quantizer.
 When the input $y$ does not satisfy $\norm{y}_2\leq \sqrt{n} M$, the
quantizer simply declares a failure, which we denote by $\bot$. Our
tracking code may use any such $(\theta, \eps)$-quantizer family. It
is typical in any construction of a gain-shape quantizer to have a
finite $M$ and $\eps>0$. Our analysis for finite $n$ will apply to any
such $(\theta, \eps)$-quantizer family and, in particular, will
bring-out the role of the ``bias'' $\eps$. However, when establishing
our optimality result, we instantiate it using a random spherical code
to get the desired performance.

%%%%%%%%%%%%%%%%%%%%%%%%%%%%%%%%%%%%%%%%
\subsection{Description of the successive update scheme}
All the conceptual components of our scheme are ready. We use the structure of
Lemma~\ref{lem:SimpleEstimate} and focus only on updating the estimates of the latest observed sample $X_{ks}$ at the decoder. Our
encoder successively updates the estimate of the latest
sample at the decoder by quantizing and sending estimates for errors
$Y_t$.

As discussed earlier, we must decide if we prefer a fast but loose approach or a slow but accurate approach for sending error estimates. To carefully examine this tradeoff, we opt for a more  general scheme where the $nRs$ bits 
available between two samples 
are divided into $m = s/p$ sub-fragments of length $nRp$ bits each.
We use an $nRp$ bit quantizer to refine error estimates for the latest sample $X_{ks}$ (obtained at time $t=ks$)
every $p$ slots, and send the resulting quantizer codewords 
as partial tracking codewords $(C_{ks,jp}, ..., C_{ks,(j+1)p-1})$, $0\le j \leq m-1$.
Specifically, the $k$th codeword transmission interval $I_k$ is divided 
into $m$ sub-fragments $I_{k,j}$, $1\leq j \leq m$ given by
\EQ{
I_{k,j} \triangleq \set{ks+jp, \dots, ks+(j+1)p-1},~0\leq j \leq m-1,
} 
and $(C_{ks,jp}, ..., C_{ks,(j+1)p-1})$ is transmitted in communication slots in $I_{k,j}$. 

At time instant $t=ks+jp+1$ the decoder receives the $j$th sub-fragment $(C_{ks,t-ks}, t \in I_{k,j})$ of $nRp$ bits, and uses it to refine the estimate of the latest source sample $X_{ks}$.  
Note that the fast but loose and the slow but accurate regimes described above correspond to $p = 1$ and $p = s$, respectively. In the middle of the interval $I_{k,j}$, the decoder ignores the partially received quantization code and 
retains the estimate $\hat{X}_{ks}$ of $X_{ks}$ formed at time $ks+(j-1)p+1$. It forms an estimate of the current state $X_{ks+i}$ by simply scaling $\hat{X}_{ks}$ by a factor of $\alpha^i$, as suggested by Lemma~\ref{lem:SimpleEstimate}.

Finally, we impose one more additional simplification on the decoder structure. Instead of using MMSE estimates for the latest sample, we simply update the
estimate by adding to it the quantized value of the error. Thus, the decoder has a simple linear structure.

We can use any $nRp$ bit quantizer\footnote{With an abuse of notation, we will use $Q_p$ instead of $Q_{Rp}$ to denote an $nRp$ bit quantizer.} $Q_p$ for the $n$-dimensional error vector, whereby this scheme can be easily  implemented in practice if $Q_p$ can be implemented. For instance, we can use any standard gain-shape quantizer. The performance of most quantizers can be analyzed explicitly to render them a $(\theta, \eps)$-quantizer family for an appropriate $M$ and function $\theta$. 
Later, when analyzing the scheme, we will consider a $Q_p$ coming from a $(\theta, \eps)$-quantizer family and present a theoretically sound guideline for choosing $p$. 
%%%

Recall that we denote the estimate of $X_u$ formed at the decoder at time $t \ge u$ by {$\hat{X}_{u|t}$}. 
We start by initializing $\hat{X}_{0|0}=0$ and then proceed using the encoder and the decoder algorithms outlined above. 
Note that our quantizer $Q_p$ may declare failure symbol $\bot$, in which case the decoder must still yield a nominal estimate. We will simply declare the estimate as\footnote{In analysis, we account for all these events as error. Only the probability of failure will determine the contribution of this part to the MSE since the process is mean-square bounded.} $0$ once a failure happens.

%% In the event of encoder success defined in equation \eqref{eq:encoder success}, the decoder estimates the last source symbol $X_{ks}$ based on the codewords $(C_0, \dots, C_{(k-1)s})$ and sub-fragments $(C_{ks,t-1-ks}: t \in I_{k,i})_{i \le j-1}$.  
%% This estimate is denoted by 
%% Note that, this estimate is useful only if the quantizer has not failed until time $ks+jp$.
%% If there is an encoder failure, it is indicated by sending a special sequence, $\bot$.
%where $\hat{X}_{ks \vert ks+jp}$ is the estimate of $X_{ks}$ based on the information received until time $t$. 
%%We use the $(\theta, \eps)$ quantizer for encoding with rate $Rp$ and $\theta(R) < 1$ 
%and $\theta(R) = 2^{-2R^{\prime}}$ where, $R^{\prime} = R - \eta - \gamma$ for some $\gamma > 0$ and $\eta > \ceil{\log(M/\eps)}/n$.
%% and further details of the quantization scheme $g: \R^n \to \set{0,1}^{nRp}$ used is discussed in Section~\ref{subsec:Quantization}. 
%for $t =ks+jp$ for $j \in \set{0, \dots, m-1}$ at encoder and $t \in I_{k,j+1}+1$ for $j \in [m]$ at the decoder.

We give a formal description of our encoder and decoder algorithms below. 

\noindent {\bf The encoder.} 

\begin{enumerate}
\item[1] Initialize $k=0$, $j=0$, $\hat{X}_{0|0}=0$.
\item[2] At time $t = ks + jp$,  use the decoder algorithm (to be described below) to form the estimate $\hat{X}_{ks \vert t}$ and compute the error
\EQN{
\label{eqn:PSUError}
Y_{k,j} \triangleq X_{ks} - \hat{X}_{ks \vert t},
}
where we use the latest sample $X_{ks}$ available at time $t=ks+jp$.

\item[3] Quantize $Y_{k,j}$ to $nRp$ bit as $Q_p(Y_{k,j})$. 

\item[4] If quantize failure occurs and $Q_p(Y_{k,j})=\bot$, send $\bot$ to the receiver and terminate the encoder. 

\item[5] Else, send a binary representation of $Q_p(Y_{k,j})$ as the communication 
$(C_{ks,0}, ..., C_{ks, p-1})$ to the receiver over the next $p$ communication slots\footnote{For simplicity, we do not account for the extra message symbol needed for sending $\bot$.}. 

\item[6] If $j<m-1$, increase $j$ by $1$; else set $j=0$ and increase $k$ by $1$. Go to Step 2.
\end{enumerate}

\noindent {\bf The decoder.}

\begin{enumerate}
\item[1] Initialize $k=0$, $j=0$, $\hat{X}_{0|0}=0$.
\item[2] At time $t = ks + jp$, if encoding failure has not occurred until time $t$, 
compute
%% \footnote{Note that the communication received at any time $t \in I_k$ is a function only of process samples $(X_0, \dots, X_{ks})$.} 
\EQN{
\nonumber
\hat{X}_{ks \vert ks+jp} = \hat{X}_{ks\vert ks + (j-1)p}+ Q_p(Y_{k,j-1}), 
}
and output $\hat{X}_{t|t} = \alpha^{t-ks}\hat{X}_{ks|t}$.

\item[3] Else, if encoding failure has occurred and the $\bot$ symbol is received
declare $\hat{X}_{s \vert t} = 0$ for all subsequent time instants $s \ge t$.

\item[4] At time $t = ks+jp+i$, for $i \in [p-1]$, 
 output\footnote{We ignore the partial quantizer codewords received as
$(C_{ks,jp+1}, C_{ks,jp+2}, \dots, C_{ks, jp+i-1})$ till time $t$.} $\hat{X}_{t|t} = \alpha^{t-ks}\hat{X}_{ks|ks+jp}$. 
\item[5] If $j<m-1$, increase $j$ by $1$; else set $j=0$ and increase $k$ by $1$. Go to Step 2.
\end{enumerate}

%% file: results.tex
%%%%%%%%%%%%%%%%%%%%%%%%%%%%%%%%%%%%%%%%
\section{Main Results} 
\label{sec:main_res}
%%%%%%%%%%%%%%%%%%%%%%%%%%%%%%%%%%%%%%%% 
We present results in two categories. First, we provide an explicit formula for the asymptotic maxmin tracking accuracy $\delta^*(R,s, \mathbb{X})$. Next, we present a theoretically-founded guideline for selecting a good $p$ for the successive update scheme with a $(\theta, \eps)$-quantizer family. Interestingly, the optimal choice may differ from the asymptotically optimal choice of $p=1$.

%%%%%%%%%%%%%%%%%%%%%%%%%%%%%%%%%%%%%%%%
\subsection{Characterization of the maxmin tracking accuracy}
To describe our result, we define a functions $\delta_0: \R_+ \to [0,1]$ 
and  $g: \R_+ \to [0,1]$ as 

\begin{align}
\delta_0(R) &\triangleq
  \frac{\alpha^2(1-2^{-2R})}{(1-\alpha^{2}2^{-2R})}, \text{ for all } R > 0;
\nonumber
\\
g(s) &\triangleq\frac{(1-\alpha^{2s})}{s(1-\alpha^2)},\text{ for all }s>0.
\label{eqn:G}
\end{align}
Note that $g(s)$ is a decreasing function of $s$ with $g(1)=1$. The 
result below shows that, for an appropriate choice of the quantizer, our successive update scheme with $p=1$ (the fast but loose version) 
 achieves an accuracy of $\delta_0(R)g(s)$ asymptotically, universally for all processes in $\mathbb{X}$.
\begin{thm}[Lower bound for maxmin tracking accuracy: The achievability]
\label{thm:samplingrate_ub}
For $R>0$ and $s\in \N$, the asymptotic maxmin tacking accuracy is bounded below as
\[
\delta^*(R, s, \mathbb{X})\geq \delta_0(R)g(s).
\]
Furthermore, this bound can be obtained by a successive update scheme with $p=1$ and appropriately chosen quantizer $Q_p$.
\end{thm}
We provide a proof in Section \ref{sec:asymptotic_achievability}. Note that while we assume that the per dimension fourth moment of the processes in $\mathbb{X}$ is bounded, the {asymptotic} result above does not depend on that bound. Interestingly, the performance characterized above is the best possible.

\begin{thm}[Upper bound for maxmin tracking accuracy: The converse]
\label{thm:samplingrate_lb}
For $R>0$ and $s\in \N$, the asymptotic maxmin tacking accuracy is bounded above as
\[
\delta^*(R, s, \mathbb{X})\leq \delta_0(R)g(s).
\]
Furthermore, the upper bound is obtained by considering a Gauss-Markov process.
\end{thm}
We provide a proof in Section \ref{sec:converse}. Thus, $\delta^*(R, s, \mathbb{X})= \delta_0(R)g(s)$ with the {\em fast but loose} successive update scheme
being universally (asymptotically) optimal and the Gauss-Markov process being the most difficult process to track. Clearly, the best possible choice of sampling period is $s=1$ and the highest possible accuracy at rate $R$ is $\delta_0(R)$, whereby we cannot hope for an accuracy exceeding $\delta_0(R)$. 

Alternatively, the results above can be interpreted as saying that we cannot subsample at a frequency less than $1/\lfloor
g^{-1}(\delta/\delta_0(R))\rfloor$ for attaining a tracking accuracy $\delta\le\delta_0(R)$. 

%%%%%%%%%%%%%%%%%%%%%%%%%%%%%%%%%%%%%%%%
\subsection{Guidelines for choosing a good $p$}
The proof of Theorem~\ref{thm:samplingrate_ub} entails the analysis of
the successive update scheme for $p=1$. In fact, we can analyze this
scheme for any $p\in\N$ and for any $(\theta, \eps)$-quantizer family; we term this tracking code
the {\em $p$-successive update} ($p$-SU) scheme. 
This analysis can provide a simple guideline for the optimal choice of $p$ depending on the performance 
of the quantizer.

However, there are some technical caveats. A quantizer family will
operate only as long as the input $y$ satisfies $\norm{y}_2\leq M$. If
a $y$ outside this range is observed is observed, the quantizer will declare $\bot$ and the
tracking code encoder, in turn, will declare a failure. We denote by
$\tau$ the stopping time at which encoder failure occurs for the first
time, $i.e.$,
\[
\tau \triangleq \min\{ks+jp: Q_p(Y_{k,j})=\bot, 0\leq k, 0 \le j\le m-1\}.
\]
Further, denote by $A_t$ the event that failure does not occur until time $t$, $i.e.$,
\[
A_t \triangleq \{\tau>t\}. 
\]
We characterize the performance of a $p$-SU in terms of the probability of encoder failure in a finite time horizon $T$.
\begin{thm}[Performance of $p$-SU]\label{thm:speed-curve}
  For fixed  $\theta, \eps, \beta\in[0,1]$, 
consider the $p$-SU scheme with an $nRp$ bit $(\theta,\eps)$-quantizer
 $Q_p$, and denote
the corresponding tracking code by $(\phi_p, \psi_p)$. Suppose that for a time horizon $T\in \N$, the tracking code $(\phi_p, \psi_p)$ satisfies $\bPr{\tau\leq T}\leq \beta$. Then, 
\[
\sup_{X\in \mathbb{X}_n}\overline{D}^T(\phi_p, \psi_p, X)\leq B_T(\theta, \eps,\beta),
\]
where $B_T(\theta, \eps, \beta)$ satisfies
\eq{
\limsup_{T\to \infty} B_T(\theta, \eps,\beta)
\leq
  \sigma^2\Big[1 -
\frac{g(s)\,\alpha^{2p}}{1-\alpha^{2p}\,\theta}
\Big(1-\frac{\eps^2}{\sigma^2} - \theta\Big)
    \Big]
  +   \frac{\kappa \beta g(s)}{(1 - \alpha^{2s})}\Big(1 -
\alpha^{2(s+p)} \frac{(1-\theta)}
{1-\alpha^{2p}\theta}\Big).
}
\end{thm}
We remark that $\beta$ can be made small by choosing $M$ to be large for a quantizer family. Furthermore, the inequality in the upper bound for the MSE in the previous result (barring the dependence on $\beta$) comes from the inequality in the definition of a $(\theta, \eps)$-quantizer, rendering it a good proxy for the performance of the quantizer. The interesting regime is that of very small $\beta$ where the encoder failure doesn't occur during the time horizon of operation.
If we ignore the dependence on $\beta$, the accuracy of the $p$-SU does not depend either on $s$ or on the bound for the fourth moment $\kappa$. 
Motivated by these insights, we define the {\em accuracy-speed} curve of a quantizer family as follows.

\begin{defn}[The accuracy-speed curve]
  For  $\alpha\in[0,1]$, $\sigma^2$, and $R>0$, the {\em accuracy-speed} curve for a $(\theta,\eps)$-quantizer family $Q$ is given by
  \[
  \Gamma_Q(p) = \frac{\alpha^{2p}}{1-\alpha^{2p}\,\theta(Rp)}
\left(1-\frac{\eps^2}{\sigma^2} - \theta(Rp)\right),\quad p>0.
  \]
\end{defn}
By Theorem~\ref{thm:speed-curve}, it is easy to see that the accuracy (precisely the upper bound on the accuracy) of a $p$-SU scheme is better when $\Gamma_Q(p)$ is larger. Thus, a good choice of $p$ for a given quantizer family $Q$ is the one that maximizes $\Gamma_Q(p)$ for $1\leq p \leq s$.

We conclude by providing accuracy-speed curves for some illustrative examples. To build some heuristics, note that a uniform quantization of $[-M,M]$ has $\theta(R)=0$ and $\eps=M 2^{-R}$. For a gain-shape quantizer, we express a vector $y=\norm{y}_2 y_s$ where the shape vector $y_s$ has $\norm{y_s}_2=1$. An ideal shape quantizer (which only can be shown to exist asymptotically) using $R$ bits per dimension
will satisfy $\E\norm{\hat {y_s}- y_s}_2^2\leq 2^{-2R}$, similar to
the scalar uniform quantizer. In one of the examples below, we consider gain-shape quantizers with such an ideal shape quantizer. 

\begin{exmp}
  We begin by considering an ideal quantizer family with $\theta(R)=2^{-2R}$ and $\eps=0$. In our asymptotic analysis, we will show roughly that such a quantizer with very small $\eps$ exists. For this ideal case, for $R>0$, the accuracy-speed curve is given by
  \[
\Gamma_Q(p)=\frac{\alpha^{2p} - \alpha^{2p}\, \theta(Rp)}{1-\alpha^{2p}\,\theta(Rp)}
= 1- \frac{1- \alpha^{2p}}{1-\alpha^{2p}2^{-Rp}}.
\]
It can be seen that $\Gamma_Q(p)$ is decreasing in $p$ whereby the optimal choice of $p$ that maximized $\Gamma_Q(p)$ over $p\in [s]$ is $p=1$. Heuristically, this justifies why asymptotically the fast but loose successive update scheme is optimal.
\end{exmp}  

\begin{exmp}[Uniform scalar quantization]
  In this example, we consider a coordinate-wise uniform quantizer. Since we seek quantizers for inputs $y\in \R^n$ such that $\norm{y}_2\leq M \sqrt{n}$, we can only use uniform quantizer of $[-M \sqrt{n},M \sqrt{n}]$ for each coordinate. For this quantizer, we have $\theta=0$ and $\eps^2=nM^22^{-2R}$, whereby the accuracy-speed curve is given by $\Gamma_Q(p)=\alpha^{2p}(1-nM^22^{-2R}/\sigma^2)$. Thus, once again, the optimal choice of $p$ that maximizes accuracy is $p=1$.
\end{exmp}  

  \begin{exmp}[Gain-shape quantizer]\label{ex:gain-shape}
    Consider the quantization of a vector $y = a y_s$ where $a = \norm{y}_2$. 
    The vector $y$ is quantized by a gain-shape quantizer which quantizes the norm and shape of the vector separately to give $Q(y) = \hat{a}\hat{y}_s$.
We use a uniform quantizer within a fixed range
$[0,M\sqrt{n}]$ in order to quantize the norm $a$ to
$\hat{a}$, where an ideal shape quantizer is employed in quantizing
the shape vector $y_s$. Namely, we assume $\E\norm{y_s -
  \hat{y}_s}_2^2 \le 2^{-2R}$ and $\norm{\hat{y}_s} \le 1$.
Suppose, that we allot $\ell$ bits out of the total budget of $nR$ bits for norm quantization and the rest for shape quantization.
Then, we see that
\EQ{
\E\norm{y - Q(y)}_2^2 \le 2a^2 2^{-2(R-\ell/n)} + nM^2 2^{-2\ell-1},
}
whereby $\theta(R) = 2^{-2(R-\ell/n)+1}$ and $\epsilon^2 = M^2 2^{-2\ell-1}$.
Thus, the accuracy-speed curve is given by
\eq{
  \Gamma_Q(p) = \frac{\alpha^{2p}}{1 - 2\alpha^{2p}2^{-2(Rp - \ell/n)}}\,
 \Big( 1 - 
\frac{2M^2 2^{-2\ell-1}}{\sigma^2} - 2^{-2(Rp - \ell/n)+1}\Big).
}

%Then, we see that
%\EQ{
%\E\norm{y - Q(y)}_2^2 \le 2a^2 2^{-2(R-l)} + 2nM^2 2^{-2n\ell},
%}
%whereby $\theta(R) = 2^{-2(R-l)+1}$ and $\epsilon^2 = M^2 2^{-2\ell+1}$.
%Thus, the accuracy-speed curve is given by,
%\EQ{
%\Gamma_Q(p) = \frac{\alpha^{2p}}{1 - \alpha^{2p}2^{-2(Rp - \ell)}}\Big( 1 - 
%\frac{M^2 2^{-2\ell}}{\sigma^2} - 2^{-2(Rp - \ell)}\Big).
%}
Note that the optimal choice of $p$ in this case depends on the choice of $M$.
\end{exmp}
We illustrated application of our analysis for idealized quantizers, but it can be used to analyze even very practical quantizers, such as the recently proposed almost optimal quantizer in~\cite{MayekarTyagi19}.

%% file: analysis_SU.tex
%%%%%%%%%%%%%%%%%%%%%%%%%%%%%%%%%%%%%%%%%%%%%%%%%%%%%%%%
\section{Analysis of the Successive Update scheme}
\label{sec:analysis_su}
%%%%%%%%%%%%%%%%%%%%%%%%%%%%%%%%%%%%%%%%%%%%%%%%%%%%%%%%
From the discussion in section \ref{sec:successive_update}, we observe that the successive update scheme is designed to refine the estimate of $X_{ks}$ in each interval $\tilde{I}_k$.
This fact helps us in establishing a recursive relation for $D_t(\phi_p, \psi_p, X)$, $t \in \tilde{I}_k$ in terms of $D_{ks}(\phi_p, \psi_p, X)$ which is provided next. 
\begin{lem}
\label{lem:sa-upper bound}
For a time instant $t = ks+jp+i, 0\leq j\leq m-1, 0\leq i \leq p-1$ and $k \ge 0$, 
let $(\phi_p, \psi_p)$ denote the tracking code of a
$p$-SU scheme employing an $nRp$ bit $(\theta,\epsilon)$-quantizer. 
Assume that $\bPr{A_t^c} \le \beta^2 $. Then, we have
\eq{
 D_t(\phi_p, \psi_p, X) \le
\alpha^{2(t-ks)}\theta^{j}D_{ks}(\phi_p, \psi_p, X)+ 
 \sigma^2(1-\alpha^{2(t-ks)}) +\frac{\alpha^{2(t-ks)}(1-\theta^{j})\eps^2}{(1-\theta)} + \alpha^{2(t-ks)}\kappa\beta.
}
\end{lem}
%\begin{lem}
%\label{lem:sa-upper bound}
%For each time instant $t = ks+jp+i,~ j+1 \in [m], ~i \in [p]$ and $k \ge 0$, the $p$-SU scheme employing a $nRp$ bit $(\theta,\epsilon)$ quantizer satisfies
%\eq{
%& D_t(\phi_p, \psi_p, X) \le
%\alpha^{2(t-ks)}\theta^{j}D_{ks}(\phi_p, \psi_p, X)+ \\ 
%& \sigma^2(1-\alpha^{2(t-ks)}) +\frac{\alpha^{2(t-ks)}(1-\theta^{j})\eps^2}{(1-\theta)} + \alpha^{2(t-ks)}\kappa\beta.
%}
%where the tracking code $(\phi_p, \psi_p)$ satisfies $\bPr{A_t^c} \le \beta $.
%%for some constant $k'$.
%\end{lem}
\ieeeproof{ 
From the evolution of the AR[1] process defined in~\eqref{eqn:process}, 
we see that $X_{t} =\alpha^{t-ks} X_{ks} + \sum_{u=ks+1}^{t}\alpha^{t-u}\xi_{u}$. 
Further for the $p$-SU scheme, we know that $\hat{X}_{t \vert t} = \alpha^{t-ks} \hat{X}_{ks|ks+jp}$ at each instant $t = ks + jp +i$. 
Therefore, we have 
\EQ{
X_t - \hat{X}_{t\vert t} = \alpha^{t-ks}(X_{ks}- \hat{X}_{ks|ks+jp}) + \sum_{u=ks+1}^{t}\alpha^{t-u}\xi_{u}.
}
Since the estimate $\hat{X}_{ks|ks+jp}$ is a function of samples $(X_0, \dots, X_{ks})$, 
and the sequence $(\xi_u, u \ge ks)$ is independent of the past, 
we obtain the per dimension MSE as 
\eq{
{D_t(\phi_p, \psi_p, X)} 
= \frac{\alpha^{2(t-ks)}}{n}\E\norm{X_{ks} -
    \hat{X}_{ks|ks+jp}}_2^2+ \sigma^2(1-\alpha^{2(t-ks)}).  
}
Further, we divide the error into two terms based on occurrence of the failure event as follows:
\eqn{
\label{D_(t)_pSU}
D_t(\phi_p, \psi_p, X) =
\frac{\alpha^{2(t-ks)}}{n}\Big[\E[\norm{X_{ks}
    -\hat{X}_{ks|ks+jp}}_2^2\SetIn{A_t}] + 
\E[\norm{X_{ks}
    -\hat{X}_{ks|ks+jp}}_2^2\SetIn{A_t^c}]\Big]+
\sigma^2(1-\alpha^{2(t-ks)}).
}
Recall that at each instant $t = ks+jp$, we refine the estimate 
$\hat{X}_{ks|ks+(j-1)p}$
of
$X_{ks}$ to $\hat{X}_{ks|ks+jp} = (\hat{X}_{ks|ks+(j-1)p}+
Q_p(Y_{k,j-1}))\SetIn{A_t}$. Upon substituting this expression for $\hat{X}_{ks|ks+jp}$, we obtain
\eq{
{\E\big[\norm{X_{ks} - \hat{X}_{ks|ks+jp}}_2^2\SetIn{A_{t}}]}
&= \E[\norm{Y_{k,j-1}- Q_p(Y_{k,j-1})}^2\SetIn{A_{t}}\big]\\
\\
&\le \theta\E[\norm{X_{ks} - \hat{X}_{ks|ks+(j-1)p}}_2^2\SetIn{A_{t}}] + n\eps^2,
}
where the identity uses the  definition of error $Y_{k,j-1}$ given in~\eqref{eqn:PSUError} and the 
 inequality holds since $Q_p$ is a $(\theta,\eps)$-quantizer.
Repeating the previous step recursively, we get
\eq{
{\frac{1}{n}\E[\norm{X_{ks} - \hat{X}_{ks|ks+jp}}_2^2\SetIn{A_{t}}]} 
&\le
\theta^{j}\cdot \frac{1}{n}\E[\norm{X_{ks} - \hat{X}_{ks \vert ks}}_2^2\SetIn{A_{t}}]
+\frac{1-\theta^{j}}{1-\theta}\cdot \eps^2
\\
&\le
\theta^{j}\cdot \frac{1}{n}\E\norm{X_{ks} - \hat{X}_{ks \vert ks}}_2^2
+\frac{1-\theta^{j}}{1-\theta}\cdot \eps^2,
}
which is the same as
\eq{
{\frac{1}{n}\E[\norm{X_{ks} - \hat{X}_{ks|ks+jp}}_2^2\SetIn{A_{t}}]}
\le
\theta^{j}\cdot D_{ks}(\phi_p, \psi_p, X)+\frac{1-\theta^{j}}{1-\theta}\cdot \eps^2.
}
Moving to the error term $\E[\norm{X_{ks} -\hat{X}_{ks|ks+jp}}_2^2\SetIn{A_t^c}]$ when encoder failure occurs,
recall that the decoder sets the estimate to $0$ in the event of an encoder failure.
Thus, using the Cauchy-Schwartz inequality, we get
\eq{
{\frac{1}{n}\E[\norm{X_{ks} - \hat{X}_{ks|ks+jp}}_2^2\SetIn{A_{t}^c}] }
&= \frac{1}{n}\E[\norm{X_{ks}}^2\SetIn{A_{t}^c}]
\\
&\le \frac{1}{n} \sqrt{\bEE{\norm{X_{ks}}^4}P(A_t^c)}
\\
&\le \kappa\beta .
} 
Substituting the two bounds above in~\eqref{D_(t)_pSU}, we get the result.
}
The following recursive bound can be obtained using almost the same proof as that of
Lemma~\ref{lem:sa-upper bound}; we omit the details. 
\begin{lem}
\label{lem:recursion2}
Let $(\phi_p, \psi_p)$ denote the tracking code of a
$p$-SU scheme employing an $nRp$ bit $(\theta,\epsilon)$-quantizer. 
Assume that $\bPr{A_t^c} \le \beta^2 $. Then, we have
\eq{
{D_{ks}(\phi_p, \psi_p,\mathbb{X}_n)} 
\le \alpha^{2s}\theta^mD_{(k-1)s}(\phi_p, \psi_p,\mathbb{X}_n) + 
\sigma^2(1-\alpha^{2s}) 
+\alpha^{2s}\eps^2\frac{(1-\theta^{m})}{(1-\theta)}+ \alpha^{2s}\kappa \beta.
}
\end{lem}
We also need the following technical observation. 
\begin{lem}
\label{lem:InductiveUB}
For a sequence $(X_k \in \R: k \in \Z_+)$ that satisfies sequence of upper bounds 
\EQ{
X_k \le a X_{k-1} + b,\qquad \forall\,k \in \Z_+,
}
with constants $a,b \in \R$ such that  $b$ is finite and $a \in (-1,1)$, we have
\EQ{
\lim_{K \to \infty}\frac{1}{K}\sum_{k=0}^{K-1}X_k \le  \frac{b}{1-a}.
}
\end{lem}
\ieeeproof{
From the sequence of upper bounds, we can inductively show that 
\EQ{
X_k \le a^kX_0 + b\frac{1-a^k}{1-a},\qquad \forall\, k \in \Z_+.
}
Averaging $X_k$ over the horizon $\set{0, \dots, K-1}$, we get 
\EQ{
\frac{1}{K}\sum_{k=0}^{K-1}X_k \le \frac{1-a^K}{K(1-a)}\Big(X_0 - \frac{b}{1-a}\Big) + \frac{b}{1-a}.
}
From the finiteness of $X_0, b$ and the fact that $\abs{a} < 1$, the result follows by taking the limit $K$ growing arbitrarily large on both sides.  
}
We are now in a position to prove Theorem~\ref{thm:speed-curve}.

\emph{Proof of Theorem \ref{thm:speed-curve}:}
We begin by noting that, without any loss of generality, we can restrict to $T=Ks$. This holds since the contributions of the error term within the fixed interval $I_K$ are bounded. 
For $T=Ks$, the time duration $\set{0, \dots, T}$ can be partitioned into intervals $(I_k, k+1\in [K])$.  
Therefore, we can write the average MSE per dimension for the $p$-SU scheme for time-horizon $T= Ks$ as 
\EQ{
\overline{D}_T(\phi_p, \psi_p,X) 
= \frac{1}{Ks}\sum_{k=0}^{K-1}\sum_{j=0}^{m-1}\sum_{i = 0}^{p-1}D_{ks+jp+i}(\phi_p, \psi_p, X).
}
From the upper bound for per dimension MSE given in Lemma~\ref{lem:sa-upper bound}, we get
\eq{
{\sum_{i = 0}^{p-1}D_{ks+jp+i}(\phi_p, \psi_p, X)}
&\leq \sum_{i=0}^{p-1}\Big[\alpha^{2(jp+i)}\theta^{j}D_{ks}(\phi_p, \psi_p, X)+
   \sigma^2(1-\alpha^{2(jp+i)}) +\frac{\alpha^{2(jp+i)}(1-\theta^{j})\eps^2}{(1-\theta)} + \alpha^{2(jp+i)}\kappa\beta\Big]
\\
&=\alpha^{2jp}\frac{1-\alpha^{2p}}{1-\alpha^2}\Big(\theta^jD_{ks}(\phi_p, \psi_p, X) +\frac{(1-\theta^j)\eps^2}{(1-\theta)}+\kappa\beta-\sigma^2\Big)+ p\sigma^{2}.
}
Summing the expression above over $j\in\{0,...,m-1\}$ and $k\in \{0, ..., K-1\}$, and dividing by $T$, we get
\eq{
  {\overline{D}_T(\phi_p, \psi_p,X)}
& \le \sigma^2 + \frac{(1-\alpha^{2s})}{s(1-\alpha^2)}\Big(\kappa\beta-\sigma^2+\frac{\eps^2}{1-\theta}\Big)
+\frac{(1-\alpha^{2p})}{s(1-\alpha^2)}\frac{(1-\alpha^{2s}\theta^m)}{(1-\alpha^{2p}\theta)}\Big(\frac{1}{K}\sum_{k=0}^{K-1}D_{ks}(\phi_p, \psi_p,X) 
-\frac{\eps^2}{1-\theta}\Big).
}
It follows by Lemma~\ref{lem:recursion2} that 
\begin{align}
{\overline{D}_T(\phi_p, \psi_p,X)}
& \le \sigma^2 + \frac{(1-\alpha^{2s})}{s(1-\alpha^2)}\Big(\kappa\beta-\sigma^2+\frac{\eps^2}{1-\theta}\Big)
+\frac{(1-\alpha^{2p})}{s(1-\alpha^2)}\frac{(1-\alpha^{2s}\theta^m)}{(1-\alpha^{2p}\theta)}\Big(
\sup_{\{a_k\}_{k\geq 0}\in \mathcal{A}}
\frac{1}{K}\sum_{k=0}^{K-1}a_k
-\frac{\eps^2}{1-\theta}\Big),
\label{e:recursive-bound}
\end{align}
where the $\mathcal{A}$ denotes the set of $\{a_k\}_{k\geq 0}$ satisfying
\eq{
a_k
&\le \alpha^{2s}\theta^ma_{k-1} + 
\sigma^2(1-\alpha^{2s}) + \alpha^{2s}\kappa \beta
+\alpha^{2s}\eps^2\frac{(1-\theta^{m})}{(1-\theta)}.
}
We denote the right-side of \eqref{e:recursive-bound} by $B_T(\theta,\epsilon,\beta)$. Noting that by Lemma~\ref{lem:InductiveUB} any sequence $\{a_k\}_{k\geq 0}\in \mathcal{A}$ satisfies
\begin{align*}
\lim_{K \to \infty}\frac{1}{K}\sum_{k=0}^{K-1}a_k \le 
\frac{1}{1-\alpha^{2s}\theta^{m}}\Big(\sigma^2(1-\alpha^{2s})
+\alpha^{2s}\kappa \beta + \alpha^{2s}\eps^2\frac{(1-\theta^m)}{(1-\theta)}\Big),
\end{align*}
we get that
\begin{align*}
{\limsup_{T\to \infty }B_T(\theta,\epsilon,\beta)}
&\leq  \sigma^2\left(
1-  \frac{(1-\alpha^{2s})}
{s(1-\alpha^2)}
\cdot \frac{\alpha^{2p}(1-\theta)}
{(1-\alpha^{2p}\theta)}
\right)
+
\eps^2 \frac{(1-\alpha^{2s})}
{s(1-\alpha^2)}
\cdot \frac{\alpha^{2p}}
{(1-\alpha^{2p}\theta)}
+\kappa\beta \frac{1}
{s(1-\alpha^2)}\left(1-
\frac{\alpha^{2(s+p)}(1-\theta)}{(1-\alpha^{2p}\theta)}
\right),
\end{align*}
which completes the proof. \qed

%% file: achievability.tex
\section{Asymptotic Achievability Using Random Quantizer}
%%%%%%%%%%%%%%%%%%%%%%%%%%%%%%%%%%%%%%%%%
\label{sec:asymptotic_achievability}

With Theorem~\ref{thm:speed-curve} at our disposal, the proof of achievability  
can be completed by fixing $p=1$ and showing the existence of
appropriate quantizer. However, we need to handle the failure event,
and we address this first. The next result shows that the failure
probability depends on the quantizer only through $M$.    
\begin{lem}
\label{lem:failure_prob}
For fixed $T$ and $n$, consider the $p$-SU scheme with $p=1$ and an 
$nR$ bit $(\theta,\epsilon)$-quantizer $Q$ with dynamic range
$M$. Then, for every $\eta>0$, there exists an $M_0$ independent of
$n$ such that for all $M\geq M_0$, we get
\[
\bPr{A_T^c} \le \eta.
\]
\end{lem}
\ieeeproof{ The event $A_T$ (of encoder failure not happening until
  time $T$ for the successive update scheme) occurs when the errors
  $Y_{k,j}$ satisfies $\norm{Y_{k,j}}_2^2\leq nM^2$, for every $k\geq
  0$ and $0\leq j\leq s-1$ such that $t=ks+j\leq T$.  For brevity, we
  denote by $Y_t$ the error random variable $Y_{k,j}$ and
  $Y^{t-1}=(Y_1, ..., Y_{t-1})$. We note that
\begin{align*}
\bPr{A_T^c}&= \bPr{A_{T-1}^c}+\bPr{A_{T-1}\cap A_{T}^c}
\\ &=\bPr{A_{T-1}^c}+\bPr{A_{T-1}\cap \{\norm{Y_T}_2^2>nM^2\}}
\\ &=\bPr{A_{T-1}^c}+ \E\big[\SetIn{A_{T-1}}\bPr{\norm{Y_T}_2^2>nM^2|
    Y^{T-1}}\big] \\ &\leq \bPr{A_{T-1}^c}+
\E\Big[\SetIn{A_{T-1}}\frac{\E[\norm{Y_T}_2^2| Y^{T-1}]}{nM^2}\Big]
\\ &= \bPr{A_{T-1}^c}+ \frac{\E[\norm{Y_T}_2^2
    \SetIn{A_{T-1}}]}{nM^2}.
\end{align*}
Note that we have $Y_T=Y_{T-1}-Q(Y_{T-1})$ under $A_{T-1}$, whereby
\[
\E[\norm{Y_T}_2^2 \SetIn{A_{T-1}}]\leq \theta\cdot \E[\norm{Y_{T-1}}_2^2
  \SetIn{A_{T-1}}]+n\eps^2.
\]
Denoting by $\beta_T^2$ the probability $\bPr{A_{T}^c}$, the previous
two inequalities imply
\[
\beta_T^2\leq \beta_{T-1}^2+ \frac{\theta}{nM^2}
\E[\norm{Y_{T-1}}_2^2]+\frac{\eps^2}{M^2}.
\]
We saw earlier in the proof of Lemma~\ref{lem:sa-upper bound} that
$\E[\norm{Y_{T-1}}_2^2]/n$ depends only on the probability
$\beta_{T-1}^2$ that failure doesn't occur until time
$T-1$. Proceeding as in that proof, we get
\[
\beta_T^2\leq \beta_{T-1}^2+\frac 1{M^2}(c_1\beta_{T-1}+c_2),
\]
where $c_1$ and $c_2$ do not depend on $n$. Therefore, there exists
$M_0$ independent of $n$ such that for all $M$ exceeding $M_0$ we have
\[
\beta_T^2\leq \beta_{T-1}^2+ \eta,
\]
which completes the proof by
summing over $T$.  }

The bound above is rather loose, but it suffices
for our purpose. In particular, it says that we can choose $M$
sufficiently large to make probability of failure until time $T$ less
than any $\beta^2$, whereby Theorem~\ref{thm:speed-curve} can be
applied by designing a quantizer for this $M$. Indeed, we can use the
quantizer of unit sphere from~\cite{Wyner1967,Lapidoth1997TIT}, along
with a uniform quantizer for gain (which lies in $[-M,M]$) to get the
following performance. In fact, we will show that a deterministic
quantizer with the desired performance exists. Note that we already
considered such a quantizer in Example~\ref{ex:gain-shape}. But the
analysis there was slightly loose, and assumed the existence of an
ideal shape quantizer.

\begin{lem}\label{lem:quantizer-const}
For every $R,\eps, \gamma, M>0$, there exists an $nR$ bit
$(2^{-2(R-\gamma)}, \eps)$-quantizer with dynamic range $M$, for all
$n$ sufficiently large.
\end{lem}
\ieeeproof{ We first borrow a classic construction
  from~\cite{Wyner1967, Lapidoth1997TIT}, which gives us our desired
  shape quantizer.  Denote by $\mathbb{S}_n$ the $(n-1)$-dimensional
  unit sphere $\{y\in \R^n: \norm{y}_2=1\}$. For every $\gamma>0$ and
  $n$ sufficiently large, it was shown
  in~\cite{Wyner1967,Lapidoth1997TIT} that there exist $2^{nR}$
  vectors $\cC$ in $\mathbb{S}_n$ such that for every $y\in
  \mathbb{S}_n$ we can find $y^\prime\in \cC$ satisfying
\[
\inner{y, y^\prime} \ge \sqrt{1-2^{-2(R-\gamma)}}.
\]
Denoting $\cos \theta=\sqrt{1-2^{-2(R-\gamma)}}$, consider the shape
quantizer $Q_R(y)$ from~\cite{Lapidoth1997TIT} given by
\eq{
  Q_{R}\left(y\right) &\triangleq \cos\theta\cdot
  \argmin_{y^\prime\in\cC} \norm{y-y^\prime}_2^2 \\ &= \cos\theta
  \cdot \arg\max_{y^\prime \in \cC}\inner{y, y^\prime}, \quad
  \forall\,y\in \mathbb{S}_n.
}
Note that we shrink the length of
$y^\prime$ by a factor of $\cos\theta$, which will be seen to yield
the gain over the analysis in Example~\ref{ex:gain-shape}.

We append to this shape quantizer the uniform gain quantizer
$q_M:[0,M]\to [0,M]$, which quantizes the interval $[0,M]$ uniformly
into sub-intervals of length $\epsilon$.  Specifically,
$q_M(a)=\eps\lfloor a/\eps\rfloor$ and the corresponding index is
given by $\lfloor a/\eps \rfloor$.  We represent this index using its
$\ell\triangleq\lceil \log(M/\eps)\rceil$ bit binary representation
and denote this mapping by $\phi^M: [0,M]\to \{0,1\}^\ell$.

For every $y\in \R^n$ such that $\norm{y}_2^2\leq nM^2$, we consider
the quantizer
\[
Q(y)=\sqrt{n}\cdot q_M\left(\frac{\norm{y}_2}{\sqrt{n}}\right)\cdot Q_R\left(\frac{y}{\norm{y}_2}\right).
\]
For this
quantizer, for every $y\in \R^n$ with $\norm{y}_2^2= nB^2$ such that
$B\leq M$, we have
\eq{ \norm{y -Q(y)}^2_2 &= \norm{y}^2_2 +
  \norm{Q(y)}_2^2-2\inner{y, Q(y)}
\\
  &= nB^2 + n\hat{B}^2\cos^2\theta
  -2nB\hat{B}\cos\theta\inner{\tilde{y},Q_R(\tilde{y})}
  \\ &\leq nB^2
  + n\hat{B}^2\cos^2\theta -2nB\hat{B}\cos^2\theta
  \\ &=
  nB^2\sin^2\theta + n(B-\hat{B})^2\cos^2\theta
  \\ &\leq
  nB^2\sin^2\theta + n\eps^2\cos^2\theta
  \\ &\leq
  nB^22^{-2(R-\gamma)} + n \eps^2,
}
where the first inequality uses
the covering property of $\cC$.  Therefore, $Q$ constitutes an
$nR+\ell$ bit $2^{-2(R-\gamma), \eps}$-quantizer with dynamic range
$M$, for all $n$ sufficiently large. Note that this quantizer is a
deterministic one.  }

\emph{Proof of Theorem \ref{thm:samplingrate_ub}:} For any fixed
$\beta$ and $\eps$, we can make the probability of failure until time
$T$ less than $\beta$ by choosing $M$ sufficiently large. Further, for
any fixed $R,\gamma>0$, by Lemma~\ref{lem:quantizer-const}, we can
choose $n$ sufficiently large to get an $nR$ bit $(2^{-2(R-\gamma)},
\eps)$-quantizer for vectors $y$ with $\norm{y}_2^2\leq
nM^2$. Therefore, by Theorem~\ref{thm:speed-curve} applied for $p=1$,
we get that
\begin{align*}
\delta^*(R, s, \mathbb{X})\geq
\frac{g(s)\,\alpha^{2}}{1-\alpha^{2}2^{-2(R-\gamma)}}\,
\Big(1-\frac{\eps^2}{\sigma^2} - 2^{-2(R-\gamma)}\Big) -
\frac{\kappa \beta g(s)}{\sigma^2(1 - \alpha^{2s})}\,\Big(1 -
\alpha^{2(s+1)} \frac{1-\theta} {1-\alpha^{2}\theta}\Big).
\end{align*}
The proof is completed upon taking the limits as $\eps, \gamma$, and
$\beta$ go to $0$.\qed

%% file: converse.tex
%%%%%%%%%%%%%%%%%%%%%%%%%%%%%%%%%%
\section{Converse bound : Proof of Theorem \ref{thm:samplingrate_lb} }
\label{sec:converse}
The proof is similar to the converse proof in~\cite{Khina2017ITW}, but
now we need to handle the delay per transmission. 
We rely on the properties of {\em entropy power} of a random variable. 
Recall that for a continuous random variable 
$X$ taking values in $\R^n$, the entropy power of $X$ is given by 
\[
\cN(X)=\frac 1 {2\pi e}\cdot 2^{(2/n)\,  h(X)},
\]
where $h(X)$ is the differential entropy of $X$. 

Consider a tracking code $(\phi, \psi)$ of rate $R$ and sampling period $s$ and a process $X\in \mathbb{X}_n$.
We begin by noting that the state at time $t$ is related to the state
at time $t+i$ as
\[
X_{t+i}=\alpha^i X_t +\sum_{j=0}^{i-1}\alpha^j\xi_{t+i-j},
\]
where the noise $\sum_{j=0}^{i-1}\alpha^j\xi_{t+i-j}$ is independent of
$X_t$ (and the past states). In particular, for $t=ks+i$,
$1\le i<s$, we get
\eq{
{\bEE{\norm{X_t - \hat{X}_{t \vert t}}_2^2} }
&=
\bEE{\norm{\alpha^i X_{ks} - \hat{X}_{t\vert t}}_2^2} +
\sum_{j=0}^{i-1}\alpha^{2j}\bEE{\norm{\xi_{t-j}}_2^2} 
 \nonumber \\
 & = \alpha^{2i}\bEE{\norm{X_{ks} - \tilde{X}_t}_2^2} +
n(1-\alpha^{2i})\sigma^2.
}
where we define $\tilde{X}_t:=\alpha^{-i}\hat{X}_{t \vert t}$ and the first identity uses the orthogonality of noise added in each round from the
previous states and noise. 
Since the Gaussian distribution has the maximum differential entropy among all continuous random variables with a given variance, and the entropy power for a Gaussian random variable equals its variance, we get that
\[
\sigma^2(1-\alpha^2) \ge \cN(\xi_{t+i}).
\]
Therefore, the previous bound for tracking error yields
\eqn{
{D_{ks+i}(\phi,\psi,X)} \nonumber
&\ge \alpha^{2i}\frac{1}{n}\bEE{\norm{X_{ks} - \tilde{X}_{ks+i}}_2^2} + \frac{(1-\alpha^{2i})}{(1-\alpha^2)}\cN(\xi_{ks+i})\nonumber
\\
&= \alpha^{2i}\frac{1}{n}\bEE{\norm{X_{ks} - \tilde{X}_{ks+i}}_2^2} + \frac{(1-\alpha^{2i})}{(1-\alpha^2)}\cN(\xi_{1}),
\label{eq:D_ks recursion}
}
where the identity uses the assumption that $\xi_t$ are identically distributed for all $t$. 
Taking average of these terms for $t=0, .., T$, we get
\eq{
{\overline{D}_T(\phi, \psi, X)  }&= \frac 1 {nKs}\sum_{k=0}^{K-1}\sum_{i=ks}^{(k+1)s-1} \bEE{\norm{X_i -\hat{X}_{i \vert i}}_2^2} 
\\ 
 &\ge \frac 1 {nKs}
  \sum_{k=0}^{K-1}\sum_{i=0}^{s-1} \alpha^{2i}\bEE{\norm{X_{ks} -
      \tilde{X}_{ks+i}}_2^2}
+ \frac{\cN(\xi_1)}{(1-\alpha^2)}\left(1-\frac{(1-\alpha^{2s})}{s(1-\alpha^2)} \right).
}
Note that $\tilde{X}_{ks+i}$s act as estimates of
${X}_{ks}$ which depend on the communication received by the decoder
until time $ks+i$. We denote the communication received at time $t$ by
$C_{t-1}$, whereby $\tilde{X}_{ks+i}$ depends only on $C_1, ...,
C_{ks+i-1}$. In particular, the communication $C_{ks}, ..., C_{ks+i-1}$
was sent as a function of $X_{ks}$, the sample seen at time $t=ks$.

From here on, we proceed by invoking the ``entropy power bounds''
for the MSE terms. 
For random variables $X$ and $Y$ such that $P_{X|Y}$ has a conditional density, the conditional entropy power is given by $\cN(X|Y)= 1/(2\pi e)2^{2h(X|Y)/n}$.\footnote{The conditional differential entropy $h(X|Y)$ is given by $\bEE{h(P_{X|Y})}$.} Bounding
MSE terms by entropy power is a standard step that allows us
to track reduction in error due to a fixed amount of communication. 

We begin by using the following standard bound (see~\cite[Chapter
  10]{Cover2006}):\footnote{It follows simply by noting that  Gaussian
  maximizes differential entropy among all random variables with a
  given second moment and that $h(X)-h(X|Y)\le H(Y)=nR$.}
For a continuous random variable
$X$ and a discrete random variable $Y$ taking $\{0,1\}^{nR}$ values,
let $\hat {X}$ be any function of $Y$. Then, it holds that
\EQN{
\label{eq:std_bound}
\frac 1 n \E \|X-\hat {X}\|^2 \ge 2^{-2R}\cN(X).
}
We apply this result to $X_{ks}$ given $C^{ks-1}$ in the role of $X$ and
the communication $C_{ks}, .., C_{ks+i-1}$ in the role of $Y$. The
previous bound and Jensen's inequality yield
\[
\frac 1 n\bEE{\norm{X_{ks} - \tilde{X}_{ks+i}}_2^2}
\ge 2^{-2Ri}\E [\cN(X_{ks}|C^{ks-1})].
\]
Next, we recall the entropy power inequality ($cf.$~\cite{Cover2006}): For
independent $X_1$ and $X_2$, $\cN(X_1+X_2)\ge
\cN(X_1)+\cN(X_2)$. Noting that $X_{ks}=\alpha^sX_{(k-1)s}+ \sum_{j=0}^{s-1}\alpha^{j}\xi_{ks-j}$, where
$\{\xi_i\}$ is an \emph{iid} zero-mean random variable independent of $X_{(k-1)s}$, and that
$C^{ks-1}$ is a function of $X_1, ..., X_{(k-1)s}$, 
we get
\begin{align*}
{\cN(X_{ks}|C^{ks-1})}
&\ge \cN(\alpha^{s}X_{(k-1)s}|C^{ks-1}) +
\cN(\xi_{ks})\frac{(1-\alpha^{2s})}{(1-\alpha^2)}
\\
&= \alpha^{2s}\cN(X_{(k-1)s}|C^{ks-1}) +\cN(\xi_1)\frac{(1-\alpha^{2s})}{(1-\alpha^2)},
\end{align*}
where the previous identities utilizes the scaling property of differential
entropy. 
Upon combining the bounds given above and simplifying, we get
\eqn{
\overline{D}_T(\phi, \psi, X)
&\ge
\frac{ \alpha^{2s}(1-\alpha^{2s}2^{-2Rs})}{s(1-\alpha^22^{-2R})}\cdot
\frac{1}{K} \sum_{k=0}^{K-1}\E[\cN(X_{(k-1)s}|C^{ks-1})]
\nonumber
\\
&\hspace{3cm}+
\frac{\cN(\xi_1)}{(1-\alpha^2)} \left(1+ \frac{(1-\alpha^{2s})(1-\alpha^{2s}2^{-2Rs})}
{s(1-\alpha^{2}2^{-2R})}-\frac{(1-\alpha^{2s})}{s(1-\alpha^2)}\right).
\label{eq:D_T lb}
}
Finally, note that the terms $\cN(X_{(k-1)s}|C^{ks-1})$ are exactly the same as that considered in \cite[eqn. 11e]{Khina2017ITW} since they correspond to recovering
$X_{(k-1)s}$ using communication that can depend on it.
Therefore, a similar expression holds here, for the sampled process $\{X_{ks}: k \in \N\}$ .
Using the recursive bound for the tracking error in~\eqref{eq:D_ks recursion} and~\eqref{eq:std_bound}, we adapt the results of \cite[eqn. 11]{Khina2017ITW} for our case to obtain
\EQ{
\E[\cN(X_{(k-1)s}|C^{ks-1})] \ge d_{k-1}^*,
}
where the quantity $d_k^*$ is given by the recursion 
\EQ{
d_{k}^*=2^{-2Rs}\Big(\alpha^{2s}d_{k-1}^* + \cN(\xi_1)\frac{(1-\alpha^{2s})}{(1-\alpha^2)}\Big),
}
with $d_0^*=0$. 

The bound obtain above holds for any given process $X\in \mathbb{X}_n$. To obtain the best possible bound
we substitute $\xi_1$ to be a Gaussian random variable, since that would maximize $\cN(\xi_1)$. 
Specifically, we set $\{\xi_k\}$ to be a Gaussian random variable with zero mean and variance $\sigma^2$ to get $\cN(\xi) = \sigma^2(1-\alpha^2)$.
Thus, taking supremum over all distributions on both sides of~\eqref{eq:D_T lb}, we have 
\eq{
\sup_{X \in \mathbb{X}_n} \overline{D}_T(\phi, \psi, X) &\ge \frac{ \alpha^{2s}(1-\alpha^{2s}2^{-2Rs})}{s(1-\alpha^22^{-2R})}\cdot
 \frac 1 {K} \sum_{k=0}^{K-1}d_{k-1}^* 
+\sigma^2 \left(1+ \frac{(1-\alpha^{2s})(1-\alpha^{2s}2^{-2Rs})}{s(1-\alpha^{2}2^{-2R})}-\frac{(1-\alpha^{2s})}{s(1-\alpha^2)}\right),
}
where
\EQ{
d_{k}^*=2^{-2Rs}\big(\alpha^{2s}d_{k-1}^* +\sigma^2(1-\alpha^{2s})\big),
}
with $d_0^*=0$. For this sequence $d_k^*$, we can see that ($cf.$~\cite[Corollary 1]{Khina2017ITW})
\eq{
\limsup_{K\to\infty}\frac 1 {K} \sum_{k=0}^{K-1}
d_{k-1}^* =\lim_{K\to \infty} d_{k}^*=\frac{\sigma^2(1-\alpha^{2s})2^{-2Rs}}{(1-\alpha^{2s}2^{-2Rs})}.
}
Therefore, we have obtained
\begin{align*}
  {\limsup_{T\to\infty}\sup_{X \in \mathbb{X}_n} \overline{D}_T(\phi, \psi, X)}
&\ge \sigma^2\bigg(\frac{ (1-\alpha^{2s})\alpha^{2s}2^{-2Rs}}{s(1-\alpha^22^{-2R})}+ 1 -\frac{(1-\alpha^{2s})}{s(1-\alpha^2)}
+ \frac{(1-\alpha^{2s})(1-\alpha^{2s}2^{-2Rs})}{s(1-\alpha^{2}2^{-2R})}\bigg)
\\
&= \sigma^2\bigg(1-g(s)\delta_0(R)\bigg).
\end{align*}
As the previous bound holds for all tracking codes $(\phi, \psi)$, it follows that
$
\delta^*(R,s,\mathbb{X}) \le g(s)\delta_0(R).
$

%% file: discussion.tex
%%%%%%%%%%%%%%%%%%%%%%%%%%%%%%%%%%%%%%%%
\section{Discussion}
\label{sec:discussion}
{We restricted our treatment to an AR[1] process with uncorrelated components.
  This restriction is for clarity of presentation, and some of the results can be extended to
AR[1] processes with correlated components. In this case, the decoder will be replaced by a Kalman-like filter in the manner of~\cite{StavrouOC18}.
}
A natural extension of this work is the study of an optimum
transmission strategy for an  AR[$n$] process in the given
setting.  
In an AR[$n$] process, the strategy of refining the latest sample is
clearly not sufficient as the value of the process at any time instant
is dependent on the past $n$ samples. 
If the sampling is periodic, even the encoder does not have access to
all these $n$ samples unless we take a sample at every
instant. A viable alternative is to take  $n$ consecutive samples at
every sampling instant. However, even with this structure on the sampling
policy, it is not clear how must the information be transmitted. A
systematic analysis of this problem is an interesting area of future
research.

Another setting which is not discussed in the current work is where the
transmissions are of nonuniform rates. 
Throughout our work, we have assumed periodic sampling and
transmissions at a fixed rate.
For the scheme presented in this paper, it is easy to see from our
analysis that only the total number of bits transmitted in each
sampling interval matters, when the dimension is sufficient 
large. That is, for our scheme, even framing each packet (sent in each communication slot) using unequal number of
bits will give the same performance as that for equal packet size, if
the overall bit-budget per sampling period is fixed.
{A similar phenomenon was observed in~\cite{KhinaKKH19}, which allowed
the extension of some of their analysis to erasure channels with feedback. We remark that a similar extension is possible for some of our results, too.
  This behavior stems from the use of successive batches of bits to
successively refine the estimate of a single sample within any sampling
interval, whereby at the end of the sampling interval the error
corresponds to roughly that for a quantizer using the total number of
bits sent during the interval.  In general, a study of nonuniform
rates for describing each sample, while keeping bits per time-slot
fixed, will require us to move beyond uniform sampling. This, too, is
an interesting research direction to pursue. 
}

{Finally, we remark that the encoder structure we have imposed, wherein the error in the estimate of the latest sample is refined at each instant, is optimal only asymptotically and is justified only heuristically for fixed dimensions. Even for one dimensional  observation it is not clear if this structure is optimal. We believe that this is a question of fundamental interest which remains open. }

\section*{Acknowledgements}
The authors would like to thank Shun Watanabe for pointing to the reference~\cite{Lapidoth1997TIT}. 
%%%%%%%%%%%%%%%%%%%%%%%%%%%%%%%%%%%%%%%%%%%%%%%%%%%%%%%%%%%%%%%%%%

\bibliographystyle{IEEEtran}
\bibliography{tit-2019}

%% file: tit-2019.bbl
\begin{thebibliography}{10}
\providecommand{\url}[1]{#1}
\csname url@samestyle\endcsname
\providecommand{\newblock}{\relax}
\providecommand{\bibinfo}[2]{#2}
\providecommand{\BIBentrySTDinterwordspacing}{\spaceskip=0pt\relax}
\providecommand{\BIBentryALTinterwordstretchfactor}{4}
\providecommand{\BIBentryALTinterwordspacing}{\spaceskip=\fontdimen2\font plus
\BIBentryALTinterwordstretchfactor\fontdimen3\font minus
  \fontdimen4\font\relax}
\providecommand{\BIBforeignlanguage}[2]{{%
\expandafter\ifx\csname l@#1\endcsname\relax
\typeout{** WARNING: IEEEtran.bst: No hyphenation pattern has been}%
\typeout{** loaded for the language `#1'. Using the pattern for}%
\typeout{** the default language instead.}%
\else
\language=\csname l@#1\endcsname
\fi
#2}}
\providecommand{\BIBdecl}{\relax}
\BIBdecl

\bibitem{Wyner1967}
A.~D. Wyner, ``Random packings and coverings of the unit n-sphere,'' \emph{The
  Bell System Technical Journal}, vol.~46, no.~9, pp. 2111--2118, 1967.

\bibitem{Lapidoth1997TIT}
A.~Lapidoth, ``On the role of mismatch in rate distortion theory,'' \emph{IEEE
  Trans.\ Inf.\ Theory}, vol.~43, no.~1, pp. 38--47, 1997.

\bibitem{Witsenhausen1979}
H.~S. Witsenhausen, ``On the structure of real-time source coders,'' \emph{Bell
  System Technical Journal}, vol.~58, no.~6, pp. 1437--1451, 1979.

\bibitem{Teneketzis2006TIT}
D.~Teneketzis, ``On the structure of optimal real-time encoders and decoders in
  noisy communication,'' \emph{IEEE Trans.\ Inf.\ Theory}, vol.~52, pp.
  4017--4035, 2006.

\bibitem{Mahajan2007ITW}
A.~Mahajan and D.~Teneketzis, ``On real-time communication systems with noisy
  feedback,'' in \emph{IEEE Inf.\ Theory\ Workshop\ (ITW)}.\hskip 1em plus
  0.5em minus 0.4em\relax IEEE, 2007, pp. 283--288.

\bibitem{Walrand1983TIT}
J.~C. Walrand and P.~Varaiya, ``Optimal causal coding - decoding problems,''
  \emph{IEEE Trans.\ Inf.\ Theory}, vol.~29, pp. 814--819, 1983.

\bibitem{YukselS2012TIT}
S.~Yuksel, ``On optimal causal coding of partially observed markov sources in
  single and multiterminal settings,'' \emph{IEEE Transactions on Information
  Theory}, vol.~59, no.~1, pp. 424--437, 2012.

\bibitem{Linder2014TIT}
T.~Linder and S.~Y{\"u}ksel, ``On optimal zero-delay coding of vector markov
  sources,'' \emph{IEEE Transactions on Information Theory}, vol.~60, no.~10,
  pp. 5975--5991, 2014.

\bibitem{Wood2017TIT}
R.~G. Wood, T.~Linder, and S.~Y{\"u}ksel, ``Optimal zero delay coding of markov
  sources: Stationary and finite memory codes,'' \emph{IEEE Transactions on
  Information Theory}, vol.~63, no.~9, pp. 5968--5980, 2017.

\bibitem{Wong1997TAC}
W.~S. Wong and R.~W. Brockett, ``Systems with finite communication bandwidth
  constraints. i. state estimation problems,'' \emph{IEEE Transactions on
  Automatic Control}, vol.~42, no.~9, pp. 1294--1299, 1997.

\bibitem{NairEvans97}
G.~N. {Nair} and R.~J. {Evans}, ``State estimation via a capacity-limited
  communication channel,'' in \emph{IEEE Conference on Decision and Control},
  vol.~1, Dec 1997, pp. 866--871.

\bibitem{NairEvans98}
------, ``State estimation under bit-rate constraints,'' in \emph{IEEE
  Conference on Decision and Control}, vol.~1, Dec 1998, pp. 251--256.

\bibitem{Dokuchaev1999CDC}
N.~G. Dokuchaev and A.~V. Savkin, ``Recursive state estimation via limited
  capacity communication channels,'' in \emph{Proceedings of the 38th IEEE
  Conference on Decision and Control}, vol.~5.\hskip 1em plus 0.5em minus
  0.4em\relax IEEE, 1999, pp. 4929--4932.

\bibitem{Smith2003TAC}
S.~C. Smith and P.~Seiler, ``Estimation with lossy measurements: jump
  estimators for jump systems,'' \emph{IEEE Trans.\ Autom.\ Control}, vol.~48,
  no.~12, pp. 2163--2171, 2003.

\bibitem{Matveev2003TAC}
A.~S. Matveev and A.~V. Savkin, ``The problem of state estimation via
  asynchronous communication channels with irregular transmission times,''
  \emph{IEEE Trans.\ Autom.\ Control}, vol.~48, no.~4, pp. 670--676, 2003.

\bibitem{Lipsa2011TAC}
G.~M. Lipsa and N.~C. Martins, ``Remote state estimation with communication
  costs for first-order lti systems,'' \emph{IEEE Trans.\ Autom.\ Control},
  vol.~56, pp. 2013--2025, 2011.

\bibitem{Chakravorty2017TAC}
J.~Chakravorty and A.~Mahajan, ``Fundamental limits of remote estimation of
  autoregressive markov processes under communication constraints,'' \emph{IEEE
  Trans.\ Autom.\ Control}, vol.~62, pp. 1109--1123, 2017.

\bibitem{Basar2013TAC}
A.~Nayyar, T.~Basar, D.~Teneketzis, and V.~V. Veeravalli, ``Optimal strategies
  for communication and remote estimation with an energy harvesting sensor,''
  \emph{IEEE Trans.\ Autom.\ Control}, vol.~58, pp. 2246--2259, 2013.

\bibitem{Sun2017ISIT}
Y.~Sun, Y.~Polyanskiy, and E.~Uysal-Biyikoglu, ``Remote estimation of the
  wiener process over a channel with random delay,'' in \emph{IEEE Inter.\
  Symp.\ Inf.\ Theory\ (ISIT)}, Jun. 2017, pp. 321--325.

\bibitem{Linder2001TIT}
T.~Linder and G.~Lugosi, ``A zero-delay sequential scheme for lossy coding of
  individual sequences,'' \emph{IEEE Trans.\ Inf.\ Theory}, vol.~47, pp.
  2533--2538, 2001.

\bibitem{Weissman2002TIT}
T.~Weissman and N.~Merhav, ``On limited-delay lossy coding and filtering of
  individual sequences,'' \emph{IEEE Trans.\ Inf.\ Theory}, vol.~48, pp.
  721--732, 2002.

\bibitem{Weissman2001TIT}
T.~Weisman and N.~Merhav, ``Universal prediction of individual binary sequences
  in the presence of noise,'' \emph{IEEE Trans.\ Inf.\ Theory}, vol.~47, pp.
  2151--2173, 2001.

\bibitem{Matloub2006TIT}
S.~Matloub and T.~Weissman, ``Universal zero-delay joint source - channel
  coding,'' \emph{IEEE Trans.\ Inf.\ Theory}, vol.~52, pp. 5240--5249, 2006.

\bibitem{GorPin73}
M.~S.~P. A.~K.~Gorbunov, ``Nonanticipatory and prognostic epsilon entropies and
  message generation rates,'' \emph{Problems Inform. Transmission}, vol.~9, pp.
  184--191, 1973.

\bibitem{GorPin74}
------, ``Prognostic epsilon entropy of a gaussian message and a gaussian
  source,'' \emph{Problems Inform. Transmission}, vol.~10, pp. 93--109, 1974.

\bibitem{StavrouKC14}
P.~Stavrou, C.~K. Kourtellaris, and C.~D. Charalambous, ``Information
  nonanticipative rate distortion function and its applications,'' \emph{CoRR},
  vol. abs/1405.1593, 2014.

\bibitem{StavrouOC18}
P.~A. {Stavrou}, J.~{Østergaard}, and C.~D. {Charalambous}, ``Zero-delay rate
  distortion via filtering for vector-valued gaussian sources,'' \emph{IEEE
  Journal of Selected Topics in Signal Processing}, vol.~12, no.~5, pp.
  841--856, Oct 2018.

\bibitem{StavrouCCL18}
P.~A. Stavrou, T.~Charalambous, C.~D. Charalambous, and S.~Loyka, ``Optimal
  estimation via nonanticipative rate distortion function and applications to
  time-varying gauss--markov processes,'' \emph{SIAM Journal on Control and
  Optimization}, vol.~56, no.~5, pp. 3731--3765, 2018.

\bibitem{Berger2000TIT}
H.~Viswanathan and T.~Berger, ``Sequential coding of correlated sources,''
  \emph{IEEE Trans.\ Inf.\ Theory}, vol.~46, no.~1, pp. 236--246, 2000.

\bibitem{MaIshwar11}
N.~{Ma} and P.~{Ishwar}, ``On delayed sequential coding of correlated
  sources,'' \emph{IEEE Transactions on Information Theory}, vol.~57, no.~6,
  pp. 3763--3782, June 2011.

\bibitem{KhinaKKH19}
A.~{Khina}, V.~{Kostina}, A.~{Khisti}, and B.~{Hassibi}, ``Tracking and control
  of gauss–markov processes over packet-drop channels with acknowledgments,''
  \emph{IEEE Transactions on Control of Network Systems}, vol.~6, no.~2, June
  2019.

\bibitem{Khina2017ITW}
A.~Khina, A.~Khisti, V.~Kostina, and B.~Hassibi, ``Sequential coding of
  {G}auss-{M}arkov sources with packet erasures and feedback,'' in \emph{IEEE
  Inf.\ Theory\ Workshop\ (ITW)}, Nov. 2017, pp. 529--530.

\bibitem{KipnisReeves19}
A.~Kipnis and G.~Reeves, ``Gaussian approximation of quantization error for
  estimation from compressed data,'' in \emph{{IEEE} International Symposium on
  Information Theory, {ISIT}}, 2019, pp. 2029--2033.

\bibitem{Delchamps89}
D.~F. Delchamps, ``Extracting state information from a quantized output
  record,'' \emph{Systems and Control Letters}, vol.~13, no.~5, pp. 365--372,
  December 1989.

\bibitem{Borkar1997}
V.~S. Borkar and S.~K. Mitter, \emph{LQG Control with Communication
  Constraints}.\hskip 1em plus 0.5em minus 0.4em\relax Boston, MA: Springer US,
  1997, pp. 365--373.

\bibitem{Wong1999TAC}
W.~S. Wong and R.~W. Brockett, ``Systems with finite communication bandwidth
  constraints. ii. stabilization with limited information feedback,''
  \emph{IEEE Transactions on Automatic Control}, vol.~44, no.~5, pp.
  1049--1053, 1999.

\bibitem{Nair2000CDC}
G.~N. Nair and R.~J. Evans, ``Communication-limited stabilization of linear
  systems,'' in \emph{Proceedings of the 39th IEEE Conference on Decision and
  Control (Cat. No. 00CH37187)}, vol.~1.\hskip 1em plus 0.5em minus 0.4em\relax
  IEEE, 2000, pp. 1005--1010.

\bibitem{Liberzon2003TAC}
D.~Liberzon, ``On stabilization of linear systems with limited information,''
  \emph{IEEE Transactions on Automatic Control}, vol.~48, no.~2, pp. 304--307,
  2003.

\bibitem{You2010TAC}
K.~You and L.~Xie, ``Minimum data rate for mean square stabilization of
  discrete lti systems over lossy channels,'' \emph{IEEE Transactions on
  Automatic Control}, vol.~55, no.~10, pp. 2373--2378, 2010.

\bibitem{Yuksel2010TAC}
S.~Yuksel, ``Stochastic stabilization of noisy linear systems with fixed-rate
  limited feedback,'' \emph{IEEE Transactions on Automatic Control}, vol.~55,
  no.~12, pp. 2847--2853, 2010.

\bibitem{Yuksel2012TAC}
S.~Yuksel and S.~P. Meyn, ``Random-time, state-dependent stochastic drift for
  markov chains and application to stochastic stabilization over erasure
  channels,'' \emph{IEEE Transactions on Automatic Control}, vol.~58, no.~1,
  pp. 47--59, 2012.

\bibitem{Yuksel2012TIT}
S.~Yuksel, ``Characterization of information channels for asymptotic mean
  stationarity and stochastic stability of nonstationary/unstable linear
  systems,'' \emph{IEEE Transactions on Information Theory}, vol.~58, no.~10,
  pp. 6332--6354, 2012.

\bibitem{Yuksel2016SIAM}
------, ``Stationary and ergodic properties of stochastic nonlinear systems
  controlled over communication channels,'' \emph{SIAM Journal on Control and
  Optimization}, vol.~54, no.~5, pp. 2844--2871, 2016.

\bibitem{Poor1994}
H.~V. Poor, \emph{An Introduction to Signal Detection and Estimation (2nd
  Edition)}.\hskip 1em plus 0.5em minus 0.4em\relax Berlin, Heidelberg:
  Springer-Verlag, 1994.

\bibitem{gersho2012vector}
A.~Gersho and R.~M. Gray, \emph{Vector quantization and signal
  compression}.\hskip 1em plus 0.5em minus 0.4em\relax Springer Science \&
  Business Media, 2012, vol. 159.

\bibitem{MayekarTyagi19}
P.~Mayekar and H.~Tyagi, ``{RATQ}: {A} universal fixed-length quantizer for
  stochastic optimization,'' \emph{arXiv:1908.08200}, 2019.

\bibitem{Cover2006}
T.~Cover and J.~Thomas, \emph{Elements of Information Theory}, ser. A
  Wiley-Interscience publication.\hskip 1em plus 0.5em minus 0.4em\relax Wiley,
  2006.

\end{thebibliography}
